\documentclass[a4paper,11pt,DIV14]{scrartcl}
\usepackage{amssymb}
\usepackage{amsmath}
\usepackage{mathtools}
\usepackage{bbold}
\usepackage{slashed}
\usepackage{graphicx}
\usepackage{bbm}
\usepackage{subfigure}
\usepackage{multirow}
\usepackage{pifont}
\usepackage{authblk}
\usepackage{enumerate}
\usepackage{lscape}
\usepackage{units}
\usepackage[square,sort&compress,numbers]{natbib}
\usepackage[UKenglish]{babel}
\usepackage[UKenglish]{isodate}
\usepackage{xcolor}
\definecolor{dark-green}{rgb}{0.1,0.7,0.3}

\usepackage[colorlinks=true,linkcolor=red,urlcolor=blue,citecolor=dark-green,menucolor=black,hyperfootnotes,linktocpage]{hyperref}

\newcommand{\mee}{\langle m_{ee}\rangle}
\newcommand{\obb}{0\nu\beta\beta}
\newcommand{\mwl}{m_{W_L}}
\newcommand{\mwr}{m_{W_R}}

\newcommand{\dd}{\mathrm{d}}

\newcommand{\csection}[1]{\section*{\centering #1}}
 
\allowdisplaybreaks

\cleanlookdateon

\title{ \Large \sffamily\bfseries Sterile neutrinos and right-handed currents in KATRIN}
\author{\sf James Barry\footnote{E-mail: {\tt james.barry@mpi-hd.mpg.de}}}
\author{\sf Julian Heeck\footnote{E-mail: {\tt julian.heeck@mpi-hd.mpg.de}}}
\author{\sf Werner Rodejohann\footnote{E-mail: {\tt werner.rodejohann@mpi-hd.mpg.de}}}
\affil{Max-Planck-Institut f\"{u}r Kernphysik, \\ Saupfercheckweg 1, 69117 Heidelberg, Germany}

\date{}

\begin{document}

\maketitle

\vspace{-1cm}

\begin{abstract}

Kurie-plot experiments allow for neutrino-mass measurements based on kinematics in an almost model-independent manner. A future tritium-based KATRIN-like experiment can be sensitive to light sterile neutrinos with masses below $\unit[18]{keV}$, which are among the prime candidates for warm dark matter. Here we consider such keV neutrinos in left--right symmetric extensions, i.e.~coupled to right-handed currents, which allow for an enhanced contribution to beta decay even for small active--sterile mixing, without violating astrophysical X-ray constraints. The modified spectral shape is in principle distinguishable from the standard contribution---especially for sterile neutrino masses below $\unit[9]{keV}$, which can lead to a distinct peak. We compare the sensitivity to constraints from the LHC and neutrinoless double beta decay.

\end{abstract}

\section{Introduction}

Neutrino oscillations have established the existence of \emph{massive} neutrinos beyond any doubt, providing precise data on the two mass-squared differences $\Delta m_{21}^2 \simeq \unit[7.5\times 10^{-5}]{eV^2}$ and $|\Delta m_{31}^2| \simeq \unit[2.4\times 10^{-3}]{eV^2}$~\cite{Beringer:1900zz}. Different types of experiments are, however, necessary to measure the \emph{absolute} neutrino mass scale, e.g.~the mass of the lightest neutrino. The arguably cleanest observation can be expected from beta-decay experiments~\cite{Drexlin:2013lha}, which rely on the purely kinematic impact of neutrino masses on the electron's energy spectrum; the tiny sub-eV masses of active neutrinos are best detected at the endpoint of the electron spectrum, where neutrinos are most non-relativistic. The upcoming Karlsruhe Tritium Neutrino experiment (KATRIN) promises a sensitivity down to neutrino masses of $\unit[0.2]{eV}$ via precise measurement of the tritium beta-decay spectrum near the endpoint at $Q \simeq \unit[18.6]{keV}$~\cite{Otten:2008zz, Angrik:2005ep}. To achieve this goal, a spectrometer filters out all electrons with energies $E < Q -\mathcal{O}(\unit[10]{eV})$ in order to focus on the relevant spectral part for eV neutrino mass measurements.

KATRIN could however be upgraded to probe the \emph{full} beta spectrum with electron energies $0\leq E \leq Q$. In order to handle the significantly enhanced electron rate of about $\unit[10^{11}]{s^{-1}}$, a modified detector system is required, and a slight decrease in energy resolution is expected from the spectrometer~\cite{Mertens:2013cias}. The expected resolution of such a KATRIN-like experiment depends, of course, largely on this new detector (e.g.~differential or integral measurements), running time, etc. 
Leaving the experimental details to the experts, let us briefly outline why a measurement of the beta spectrum in a future tritium-based KATRIN-like experiment would be worthwhile. In complete analogy to active-neutrino mass searches, the full beta spectrum carries information about neutrinos with masses up to $\unit[18.6]{keV}$, belonging to so-called sterile neutrinos. These appear naturally in many models that aim to make active neutrinos massive, the most famous example being the right-handed neutrinos of the seesaw mechanism~\cite{Minkowski:1977sc, GellMann:1980vs, Yanagida:1979as, Mohapatra:1979ia}. Depending on their mass, the additional sterile neutrinos can not only provide an explanation for the baryon asymmetry of our universe via leptogenesis~\cite{Fukugita:1986hr}, a weakly mixed keV-scale sterile neutrino is even a perfect candidate for warm dark matter, ameliorating the structure-formation problems of cold dark matter (see e.g.~Ref.~\cite{Bode:2000gq}). Since the mass $m_s$ of this dark matter candidate is required to be $\mathcal{O}(\unit{keV})$ in order to satisfy astrophysical constraints, it is of obvious interest to pursue a detection at KATRIN, which probes the same mass range. The search for heavier neutrinos in beta decay has been suggested in Ref.~\cite{Shrock:1980vy}, and the expected spectral change due to an active--sterile mixing angle~$\theta_s$ can be found in in Ref.~\cite{deVega:2011xh}. 
A dedicated discussion of the sensitivity of a KATRIN-like experiment to a keV state is currently in progress, with expected reach down to~\cite{Mertens:2013cias}
\begin{align}
\sin^2 \theta_s = \epsilon \simeq 10^{-6} {\text{--}} 10^{-8} \,,
\label{eq:KATRINsensitivity}
\end{align}
for $\unit[1]{keV}\lesssim m_s \lesssim \unit[17]{keV}$. The expected reach $\epsilon$ depends on the experimental details, for example differential vs.~integral measurement, and the sterile neutrino mass, but is expected to lie in the range $10^{-6}$--$10^{-8}$, which we will use as reference values throughout this paper. The sensitivity to the sterile neutrino mass goes down for values close to $Q$ or $0$ on statistical grounds, and we will optimistically assume the range $\unit[1]{keV}\lesssim m_s \lesssim \unit[17]{keV}$ to be testable.
Note that \emph{current} limits from kink-searches in nuclear beta decay are only of order $\sin^2 \theta_s  < 10^{-3}$~\cite{Beringer:1900zz}. See also Ref.~\cite{Abdurashitov:2014vqa} for a recent evaluation of this approach, and Refs.~\cite{Herczeg:2001vk,Severijns:2006dr} for an overview of other new-physics effects in beta decay.
Lastly, the discussion here is of course not limited to the KATRIN experiment, but qualitatively applicable to any precise measurement of beta spectra. For example, the Project~8 experiment, currently in development, aims to measure the electron energy in tritium beta decay via cyclotron radiation very precisely~\cite{Monreal:2009za}. Since it is designed to measure the full beta spectrum, it can also be used in the search for keV sterile neutrinos.

The vanilla case of a keV sterile neutrino interacting only via small active--sterile mixing may, however, not be the full story. A theoretically very appealing explanation of the seesaw mechanism is given by left--right symmetric models based on the gauge group $SU(2)_L \times SU(2)_R \times U(1)_{B-L}$~\cite{Mohapatra:1974gc, Pati:1974yy, Senjanovic:1975rk, Mohapatra:1980yp, Appelquist:2003uu}. These provide a dynamical origin for parity violation and connect the sterile neutrino masses to the scale of $B-L$ breaking. The right-handed neutrinos are then strictly speaking no longer sterile, but rather part of a right-handed $SU(2)_R$ doublet, residing in a right-handed current $j_R^\mu$ coupled to a charged vector boson $W_R$ in complete analogy to the left-handed case. We will still refer to the right-handed neutrino counterparts as sterile neutrinos in the rest of this paper. 
We stress that the additional interactions, mediated for example by the new gauge boson $W_R$, still allow for the keV neutrino to be dark matter, albeit with a modified production mechanism behind it~\cite{Bezrukov:2009th} (see Ref.~\cite{Nemevsek:2012cd} for a recent evaluation of warm dark matter in low-scale left--right symmetric theories). We are in the following not concerned with the precise production mechanism, which can provide additional constraints, but only the experimental detection of warm dark matter keV sterile neutrinos at KATRIN.

The impact of additional right-handed current interactions in KATRIN has been studied for \emph{active} neutrinos in Refs.~\cite{Bonn:2007su, Stephenson:2000mw, SejersenRiis:2011sj}, but never for keV \emph{sterile} neutrinos. Our goal in this paper is to amend this glaring omission and study the impact of right-handed currents on the detection of a keV sterile neutrino in a future KATRIN-like experiment. After briefly introducing the left--right model in Sec.~\ref{sec:model}, we will show in Sec.~\ref{sec:betadecay} that the interference of left- and right-handed currents gives rise to a modified spectral shape on top of the beta spectrum, in principle distinguishable from the standard case. We estimate possible KATRIN limits on the left--right parameters in Sec.~\ref{sec:expected_limits} and collect existing limits in Sec.~\ref{sec:existing_limits}. Sec.~\ref{sec:doublebeta} provides a rough comparison of the expected limits to those from neutrinoless double beta decay. Finally, we conclude in Sec.~\ref{sec:conclusion}.

\section{Model context}
\label{sec:model}

The left--right symmetric model (LRSM) has been well-studied in the literature; we report here only those equations relevant for our purposes and refer the reader to Ref.~\cite{Barry:2013xxa} for a
more comprehensive discussion and definitions. The introduction of right-handed neutrinos with both Dirac and Majorana mass terms allows the type I+II seesaw to be naturally accommodated in the model,
so that the full neutrino mass matrix
\begin{equation}
 M_\nu \equiv \begin{pmatrix}
               M_L & M_D \\ M_D^T & M_R
              \end{pmatrix}
\end{equation}
is diagonalized by the unitary matrix
\begin{equation}
 W \equiv \begin{pmatrix}
           U & S \\ T & V
          \end{pmatrix}
\end{equation}
to $W^\dagger M_\nu W^* = {\rm diag}(m_1,m_2,m_3,M_1,M_2,M_3)$, where $m_i$ ($M_i$) are the light (heavy) neutrino masses. Here $M_D$ is the Dirac mass matrix whereas $M_L$ and $M_R$ are the
symmetric left- and right-handed Majorana mass matrices. The diagonalization equation leads to the useful relations
\begin{equation}
 M_L = U\tilde m_\nu U^T+S\tilde M_R S^T, \quad M_D = U\tilde{m}_\nu T^T + S \tilde{M}_R V^T \quad {\rm and} \quad M_R = T\tilde m_\nu T^T + V \tilde M_R V^T,
 \label{eq:md_alt}
\end{equation}
where $\tilde m_\nu \equiv {\rm diag}(m_1,m_2,m_3)$ and $\tilde M_R \equiv {\rm diag}(M_1,M_2,M_3)$. 
The unitarity conditions $W^\dagger W = \mathbb{1}_{6\times 6} = W W^\dagger$ further yield
\begin{align}
U T^\dagger + S V^\dagger = 0 = S^\dagger U + V^\dagger T \quad\text{ and } \quad U U^\dagger + S S^\dagger = \mathbb{1}_{3\times 3} = T T^\dagger + V V^\dagger \,.
\label{eq:diagonalization}
\end{align}
The mixing matrices $U$, $S$, $T$ and $V$ can be expanded as
\begin{equation} 
\begin{gathered} 
  U \equiv \left[\mathbb{1}_{3\times 3} - \frac{1}{2}M^{}_DM_R^{-1}(M^{}_DM_R^{-1})^\dagger\right] V_\nu, \quad V \equiv \left[\mathbb{1}_{3\times 3} - \frac{1}{2}(M^{}_DM_R^{-1})^\dagger M^{}_DM_R^{-1}\right]V_R,\\[1mm] 
  S \equiv M^{}_DM_R^{-1}V_R, \quad T \equiv -(M_D M_R^{-1})^\dagger V^{}_\nu\, , \label{eq:mix_matrices} 
  \end{gathered} 
\end{equation}
where the unitary matrices $V_\nu$ and $V_R$ are defined by
\begin{equation} \begin{gathered}
 m_\nu = M_L - M_DM_R^{-1}M_D^T = V_\nu\,{\rm diag}(m_1,m_2,m_3) \,V_\nu^T , \\[1mm]
M_R = V_R\,{\rm diag}(M_1,M_2,M_3)V_R^T\, .
 \end{gathered} \label{eq:mnu_MR_def}
\end{equation}
$m_\nu$ is the light neutrino mass matrix after the heavy right-handed neutrinos have been integrated out. Note that one usually assumes a discrete symmetry in addition to the gauge symmetry, which
means that (in the case of a charge conjugation symmetry)
\begin{equation}
 M^{}_D = M_D^T \quad {\rm and} \quad M_L = \frac{v_L}{v_R}M_R\, ,\label{eq:md_ml_symm_relations}
\end{equation}
in the latter case up to complex phases. $v_{L,R}$ are the vacuum expectation values of the left- and right-handed triplet, respectively (see Ref.~\cite{Barry:2013xxa} for details). In this case the left--right
mixing can be written as~\cite{Nemevsek:2012iq}
\begin{equation}
 M^{}_DM_R^{-1} = \sqrt{\frac{v_L}{v_R} \mathbb{1}_{3\times 3} -V^{}_\nu\tilde m_\nu V_\nu^T V_R^* \tilde M_R^{-1} V_R^\dagger}\, ,
 \label{eq:mdmr_sqrt}
\end{equation}
which gives eight possible solutions.

The flavor neutrino states emitted in beta decay can be expressed as
\begin{align}
 \nu'_L &= U\nu^{}_L + SN_R^c\, , \\
 \nu'_R &= T^*\nu_L^c + V^*N_R^{}\, .
 \label{eq:flavour_eigenstates}
\end{align}
which are linear combinations of the light and heavy mass eigenstates, $\nu_L$ and $N_R$, respectively. The mixing matrices $S$ and $T$ denote the left--right or light--heavy neutrino mixing, given
roughly by the ratio of Dirac and Majorana mass scales [cf. Eqs.~\eqref{eq:mix_matrices} and \eqref{eq:mdmr_sqrt}]. In this way it is possible to have heavy neutrino mass states emitted in beta decay, as long as they are lighter than the available energy (the $Q$-value for the decay).

\section{Beta decay with right-handed currents}
\label{sec:betadecay}

\subsection{Differential electron spectrum}
\begin{figure}[tp]
 \centering
 \includegraphics[width=0.4\textwidth]{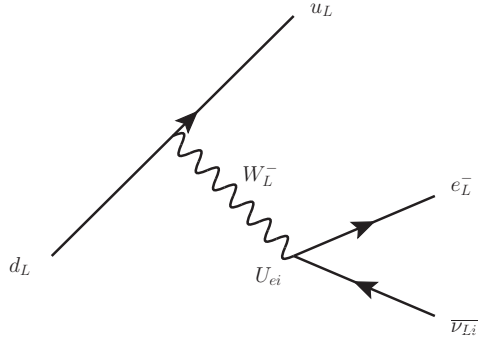}
 \caption{Feynman diagram for beta decay in the Standard Model with three massive neutrinos ($i=1,2,3$), mediated via left-handed currents.}
 \label{fig:beta_LL}
\end{figure}
The differential electron spectrum in beta decay with left-handed currents only (the standard case shown in Fig.~\ref{fig:beta_LL}) is determined by simple phase-space arguments to be
\begin{align}
\begin{split}
 \left(\frac{\dd \Gamma}{\dd E}\right)_{\rm std} &= K'\sum_i|U_{ei}|^2 (E+m_e) p_e E_{\nu_i} p_{\nu_i} \\ & = K'(E+m_e) p_e (E_0-E) \sum_i|U_{ei}|^2\sqrt{(E_0-E)^2- m_i^2}\,\Theta(E_0-E-m_i) \,,
 \label{eq:diff_spectrum_standard}
 \end{split}
\end{align}
where light antineutrinos of mass $m_i$, mixing matrix $U$, and energy $E_{\nu_i} = E_0-E$ are emitted in the process. Here, $E$ is the \emph{kinetic energy} of the electron with total energy $E+m_e$ and momentum $p_e = \sqrt{E^2+2Em_e}$, and $E_0 \equiv {\rm max}(E)$ is the endpoint energy. $K'$ includes the Fermi function $F(Z,E)$ as well as the nuclear matrix elements and relevant prefactors, i.e.,
\begin{equation}
 K' = \frac{G_F^2\cos^2\theta_C}{2\pi^3}|{\cal M}|^2 F(Z,E),
\end{equation}
where $\theta_C$ is the Cabibbo angle and ${\cal M}$ is the dimensionless nuclear matrix element~\cite{Drexlin:2013lha}. The Heaviside step function $\Theta(E_0-E-m_i)$ ensures that only physically allowed regions of the spectrum are considered. Since the recoil energy of the nucleus does not change much in the region of interest one can apply a constant recoil energy correction $E_{\rm rec}$ and define $E_0 = Q -
E_{\rm rec}$, where $Q$ is the energy released in the decay. We can divide out the factors that are independent of neutrino parameters by defining the Kurie function
\begin{equation}
 K(E) \equiv \sqrt{\frac{\dd \Gamma/\dd E}{K'(E+m_e)p_e}}\, ,
 \label{eq:standardkurie}
\end{equation}
which is linear in energy for massless neutrinos, i.e., $K(E)|_{m_i=0} = E_0-E$. Nonzero neutrino masses modify this behavior and result in a premature drop-off of $K(E)$; knowing that KATRIN is only sensitive to the quasi-degenerate neutrino mass regime $m_1 \simeq m_2 \simeq m_3 \gtrsim \unit[0.2]{eV}$, one can pool all three neutrino contributions into one parameter, the effective electron-neutrino mass in beta decay~\cite{Farzan:2001cj,Farzan:2002zq},
\begin{equation}
 m_\beta^2 \equiv \sum_i|U_{ei}|^2 m_i^2 \simeq m_1^2 \,. \label{eq:mbeta_def}
\end{equation}
The last relation follows from the unitarity of $U$ in the three-flavor scheme. In this limit, all three active-neutrino kinks fall together, resulting in the modified Kurie function
\begin{equation}
 K(E) \simeq (E_0 - E)\sqrt[4]{1-\frac{m_\beta^2}{(E_0-E)^2}}\, \Theta(E_0-E-m_\beta)\, .
\end{equation}
A precise measurement of the electron's energy spectrum close to the endpoint $E_0$ is hence sensitive to the neutrino mass parameter $m_\beta$.\footnote{The quasi-degenerate neutrino mass regime to be probed by KATRIN implies that the sum of neutrino masses $\sum_j m_j$ is larger than $\unit[0.6]{eV}$, seemingly already excluded by cosmological measurements~\cite{Ade:2013zuv}. The latter are however highly dependent on the combination of datasets and correlated with other parameters, so KATRIN's region of interest is far from excluded and will provide important complementary information.}

\begin{figure}[tp]
 \centering
 \subfigure[]{\label{fig:beta_RL}
 \includegraphics[width=0.3\textwidth]{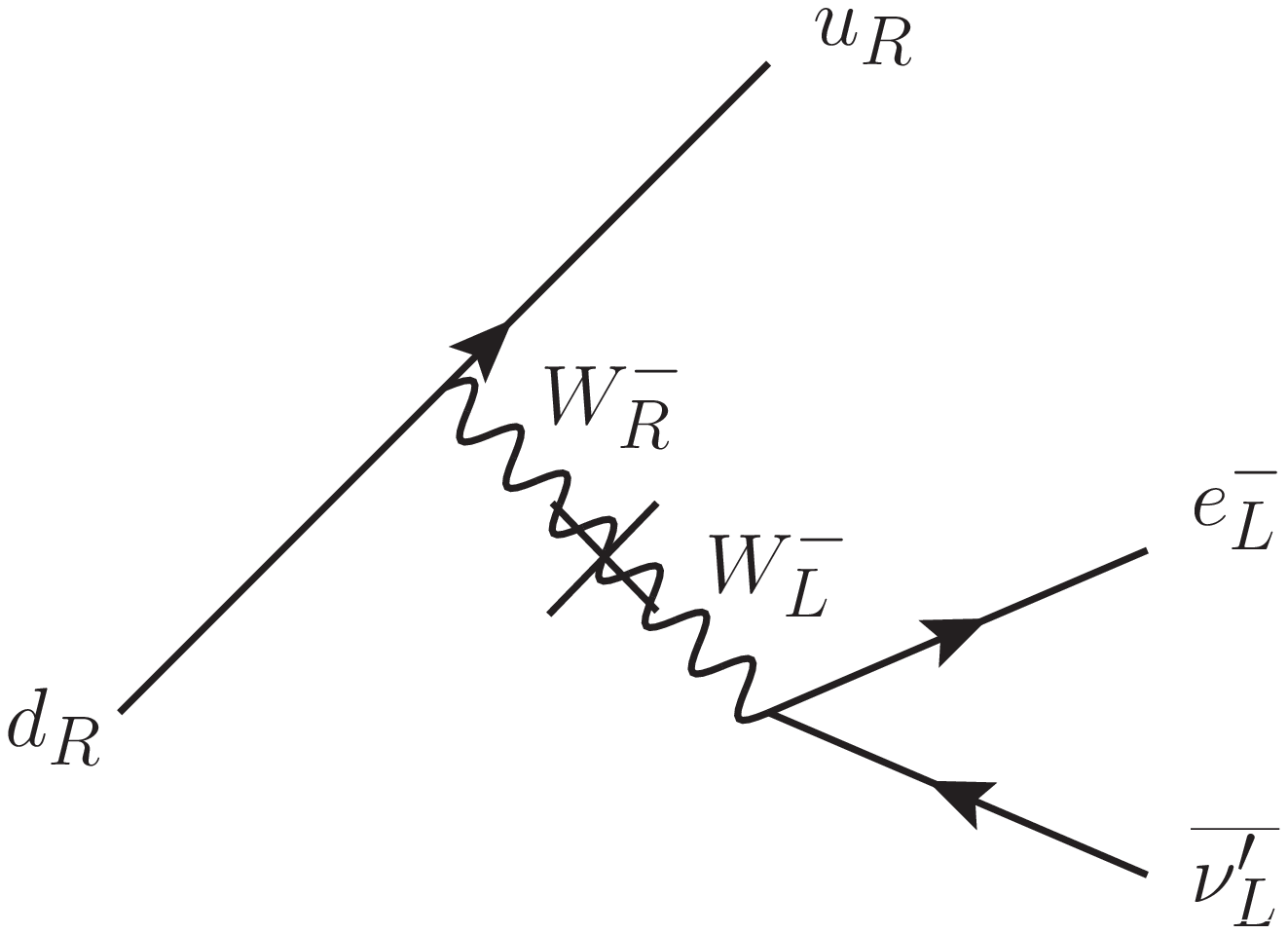}}
 \subfigure[]{\label{fig:beta_LR}
 \includegraphics[width=0.3\textwidth]{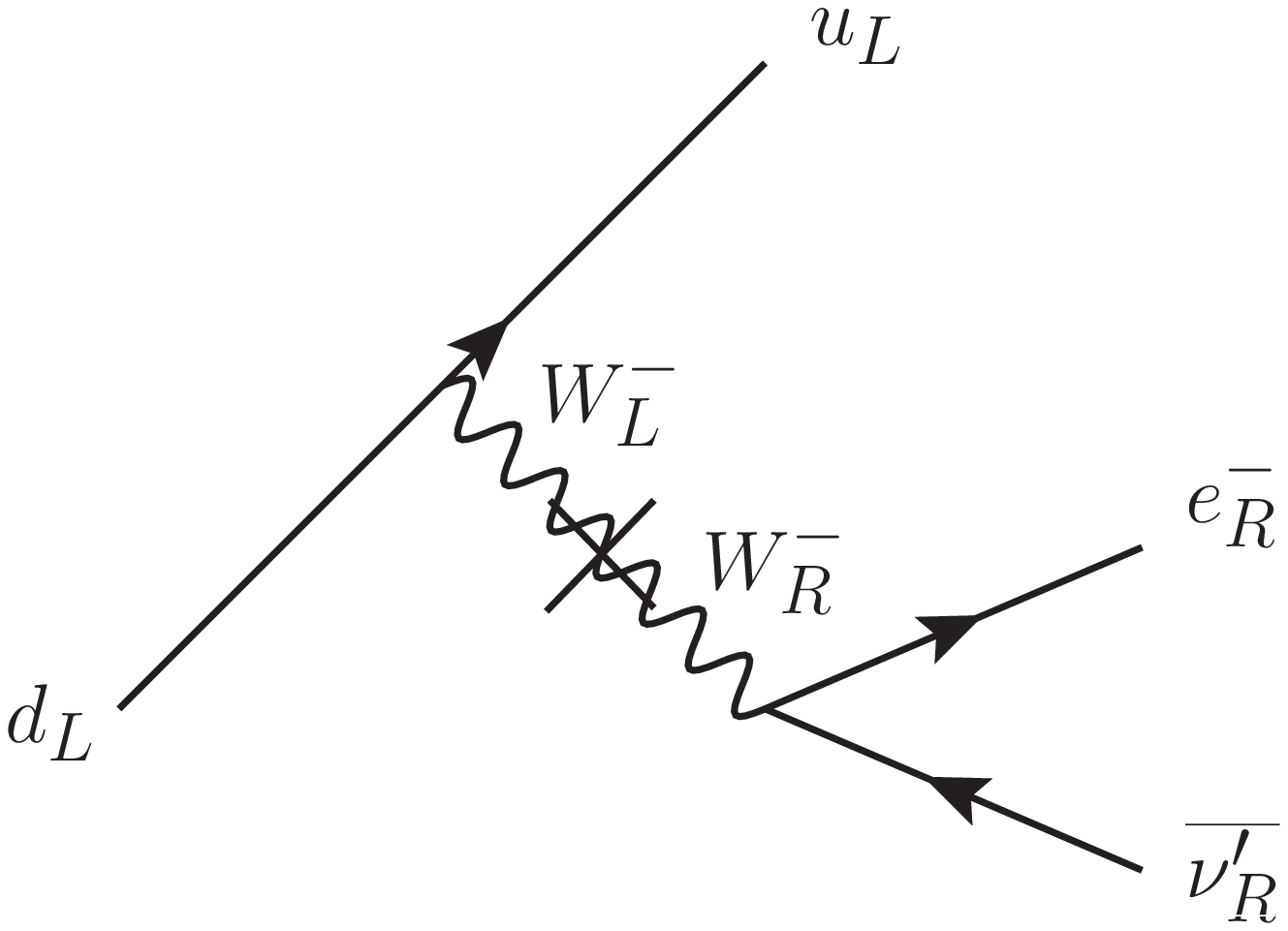}}
 \subfigure[]{\label{fig:beta_RR}
 \includegraphics[width=0.3\textwidth]{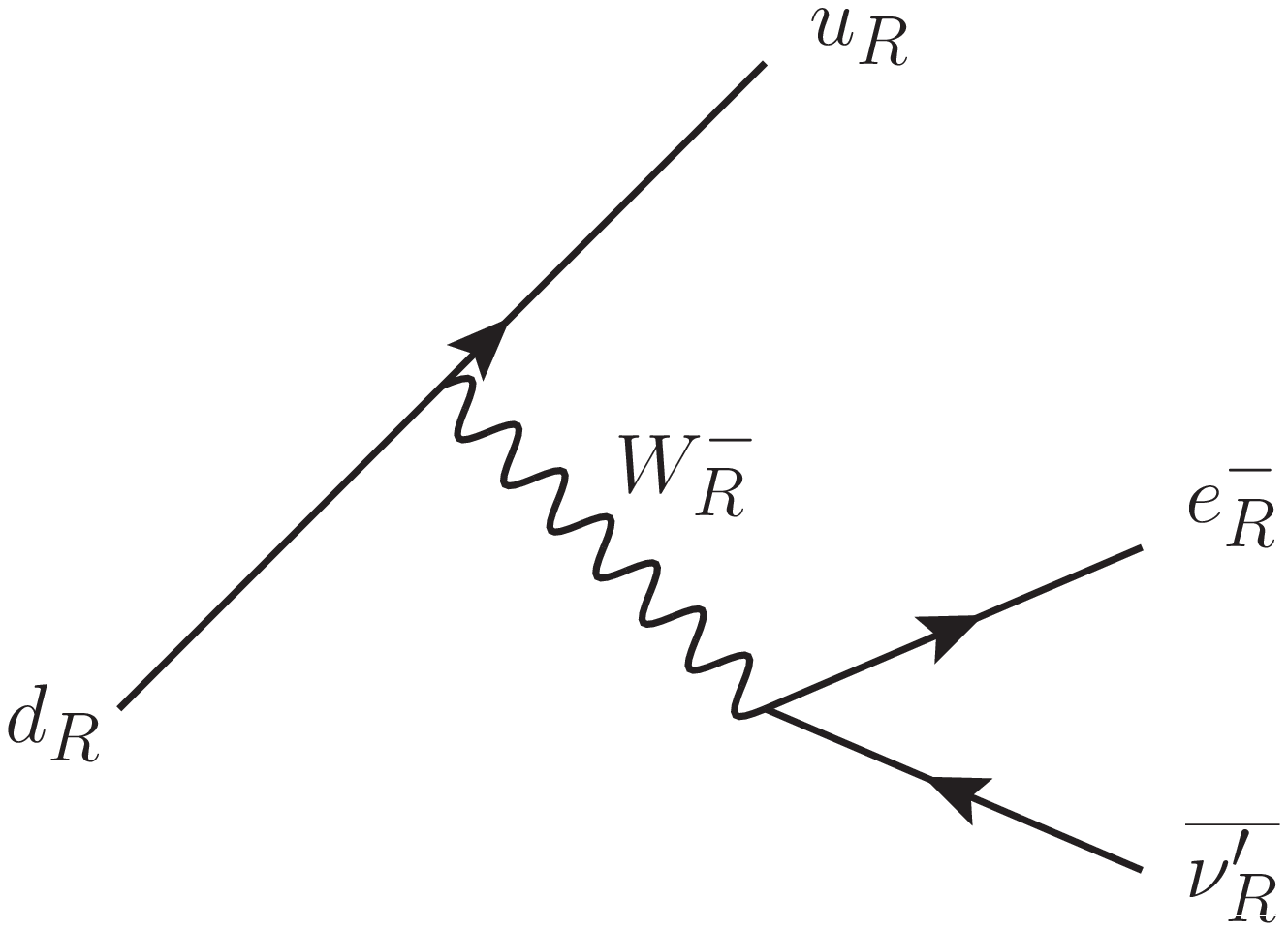}}
 \caption{Beta decay channels in the LRSM, with left- and right-handed currents at either or both fermion vertices. We use the flavor eigenstates $\nu'_{L,R}$, which are
superpositions of the mass eigenstates $\nu_L$ and $N_R$ [see Eq.~\eqref{eq:flavour_eigenstates}], so that the final result contains different combinations of the mixing matrices $U$, $S$, $T$ and
$V$.}
\label{fig:beta_LRSM}
\end{figure}
In the left--right symmetric model with equal gauge coupling constants ($g_L = g_R = g$), the process can also proceed via right-handed currents with the exchange of right-handed gauge bosons, $W_R$,
as well as via $W_L$--$W_R$ mixing. The three possible diagrams are shown in Fig.~\ref{fig:beta_LRSM}; one also has a modified version of the diagram in Fig.~\ref{fig:beta_LL} with the replacements
$\nu_L \leftrightarrow N_R^c$ and $U_{ei} \leftrightarrow S_{ei}$. 
Diagrams with left-handed lepton vertices [Figs.~\ref{fig:beta_LL} and \ref{fig:beta_RL}] lead to the differential spectrum
\begin{subequations}
\begin{align}
 \left(\frac{\dd \Gamma}{\dd E}\right)_{LL} &= K'(E+m_e) p_e (E_0-E)[1+2C \tan\xi\, \cos \alpha] \label{eq:diff_spectrum_ll}\\
  &\quad \times \sum_i\left[|U_{ei}|^2\sqrt{(E_0-E)^2 - m_i^2}\,\Theta(E_0-E-m_i) \right. \label{eq:diff_spectrum_ll_light}\\
  &\quad\quad\quad\quad \left. + |S_{ei}|^2\sqrt{(E_0-E)^2 - M_i^2}\,\Theta(E_0-E-M_i)\right], \label{eq:diff_spectrum_ll_heavy}
\end{align}
\end{subequations}
where $C=(g_V^2-3g_A^2)/(g_V^2+3g_A^2) \simeq -0.65$. The second term proportional to $\tan\xi$ in the square brackets in Eq.~\eqref{eq:diff_spectrum_ll} comes from the interference between 
Figs.~\ref{fig:beta_LL} and \ref{fig:beta_RL}. In this case the new physics diagram still has a left-handed lepton vertex, whereas the hadronic side is right-handed. The two different terms in
Eqs.~\eqref{eq:diff_spectrum_ll_light} and \eqref{eq:diff_spectrum_ll_heavy} correspond to the production of light or heavy neutrino mass eigenstates, respectively. The heavy neutrinos of mass
$M_i$ will only be emitted if $M_i < E_0$, so that for tritium beta decay only right-handed neutrinos at the keV-scale or below can be produced. Note also that $U$ is no longer unitary [cf.~Eq.~\eqref{eq:diagonalization}] but we rather have
\begin{align}
 \sum_i |U_{e i}|^2 = 1 - \sum_i |S_{e i}|^2 \,,
\end{align}
so the non-unitarity is of the same order as the sterile neutrino contribution but of opposite sign.

There will also be contributions from interference between the standard diagram and those in Figs.~\ref{fig:beta_LR} and \ref{fig:beta_RR}, i.e.,
\begin{subequations}
\begin{align}
 \left(\frac{\dd \Gamma}{\dd E}\right)_{LR} &= -2K'm_e p_e{\rm Re}\left\{\left[\left(\frac{\mwl}{\mwr}\right)^2 + C\tan\xi\,e^{i\alpha}\right] \right. \label{eq:diff_spectrum_lr}\\[1.5mm]
  &\quad \left. \times \sum_i\left[U_{ei}T_{ei}m_i \sqrt{(E_0-E)^2-m_i^2}\,\Theta(E_0-E-m_i) \right. \right. \label{eq:diff_spectrum_lr_light}\\
  &\quad \quad \quad \quad \left. \left.  + S_{ei}V_{ei}M_i\sqrt{(E_0-E)^2-M_i^2}\,\Theta(E_0-E-M_i)\right]\right\} .  \label{eq:diff_spectrum_lr_heavy}
\end{align}
\end{subequations}
Here the new physics diagram has a right-handed lepton vertex, which leads to a different mass and energy dependence from the contraction of lepton traces. The two prefactors in square brackets in
Eq.~\eqref{eq:diff_spectrum_lr} come from Figs.~\ref{fig:beta_RR} and \ref{fig:beta_LR}, respectively. Since we have $\xi <0 $ in the heavy $W_R$ limit we are interested in, and furthermore $|\tan \xi|\leq m_{W_L}^2/m_{W_R}^2$~\cite{Barry:2013xxa} and $C \simeq - 0.65$, the prefactor in Eq.~\eqref{eq:diff_spectrum_lr} will be largest in magnitude for a vanishing CP phase, $\alpha = 0$. Overall, $\left(\dd \Gamma/\dd E\right)_{LR}$ can, of course, have either sign due to the various mixing-matrix elements present in this interference term.

Once again one can have both light [Eq.~\eqref{eq:diff_spectrum_lr_light}] and heavy
[Eq.~\eqref{eq:diff_spectrum_lr_heavy}] neutrino mass eigenstates in the final state. As a consequence of left--right interference, fermion mass flips are required for both the electron and the neutrinos, effectively replacing $E+m_e$ by $m_e$ and $E_0-E$ by $m_i$ or $M_i$ in the LL formula in Eq.~\eqref{eq:diff_spectrum_ll}. Since the neutrinos are Majorana particles, these mass flips are sensitive to the Majorana phases contained in, e.g., $U$, as can be seen by a closer examination of $\left(\dd \Gamma/\dd E\right)_{LR}$. This is contrary to the standard case in which the spectrum is proportional to $|U_{ei}|^2$, independent of Majorana phases and hence insensitive to the Dirac vs.~Majorana nature of neutrinos.

As an aside, it is interesting to note that if one expands the two square roots in Eqs.~\eqref{eq:diff_spectrum_lr_light} and \eqref{eq:diff_spectrum_lr_heavy} in the limit $E_0 -E \gg M_j, m_j$, corresponding to the region of parameter and phase space where all emitted neutrinos are highly relativistic, the leading terms can be combined into one effective mass parameter
\begin{equation}
 \sum_i U_{ei} m_i T_{ei} +\sum_i^{\rm light} S_{ei} M_i V_{ei} ,
\end{equation}
similar to $m_\beta$ in Eq.~\eqref{eq:mbeta_def}.
In case \emph{all} right-handed neutrinos are lighter than $E_0$, this effective neutrino-mass parameter is simply $[M_D]_{ee}$ [using Eq.~\eqref{eq:md_alt}], reminiscent of the neutrino mass parameter relevant for neutrinoless double beta decay $[M_\nu]_{ee}$. In that sense these terms probe the $(1,1)$ element of the Dirac mass matrix, which comes from the fact that one needs a helicity flip on the external neutrino line in
order for the left- and right-handed diagrams to interfere in the first place.\footnote{The next term in the expansion, proportional to neutrino mass cubed, can in the same way be written as
$\left[|M^{}_L|^2M^{}_D +M^{}_DM_D^\dagger M^{}_D+M^{}_LM_D^*M^{}_R+M^{}_D|M^{}_R|^2\right]_{ee}$.} In the same way, the dependence on the electron mass in Eq.~\eqref{eq:diff_spectrum_lr} comes from
the helicity flip on the electron line. This is punishment for the mass flip, but very mild due to the borderline non-relativistic nature of the electrons in the region of interest for keV sterile neutrinos ($E/m_e \lesssim 18.6/511 \simeq 0.04$). The mass flip is only dangerous for relativistic particles, i.e.~the very light active neutrinos, whose contribution can be dropped in Eq.~\eqref{eq:diff_spectrum_lr_light} (seeing as it is further suppressed by $T_{e i} \mwl^2/\mwr^2$).

\begin{figure}
 \centering
 \includegraphics[width=0.48\textwidth]{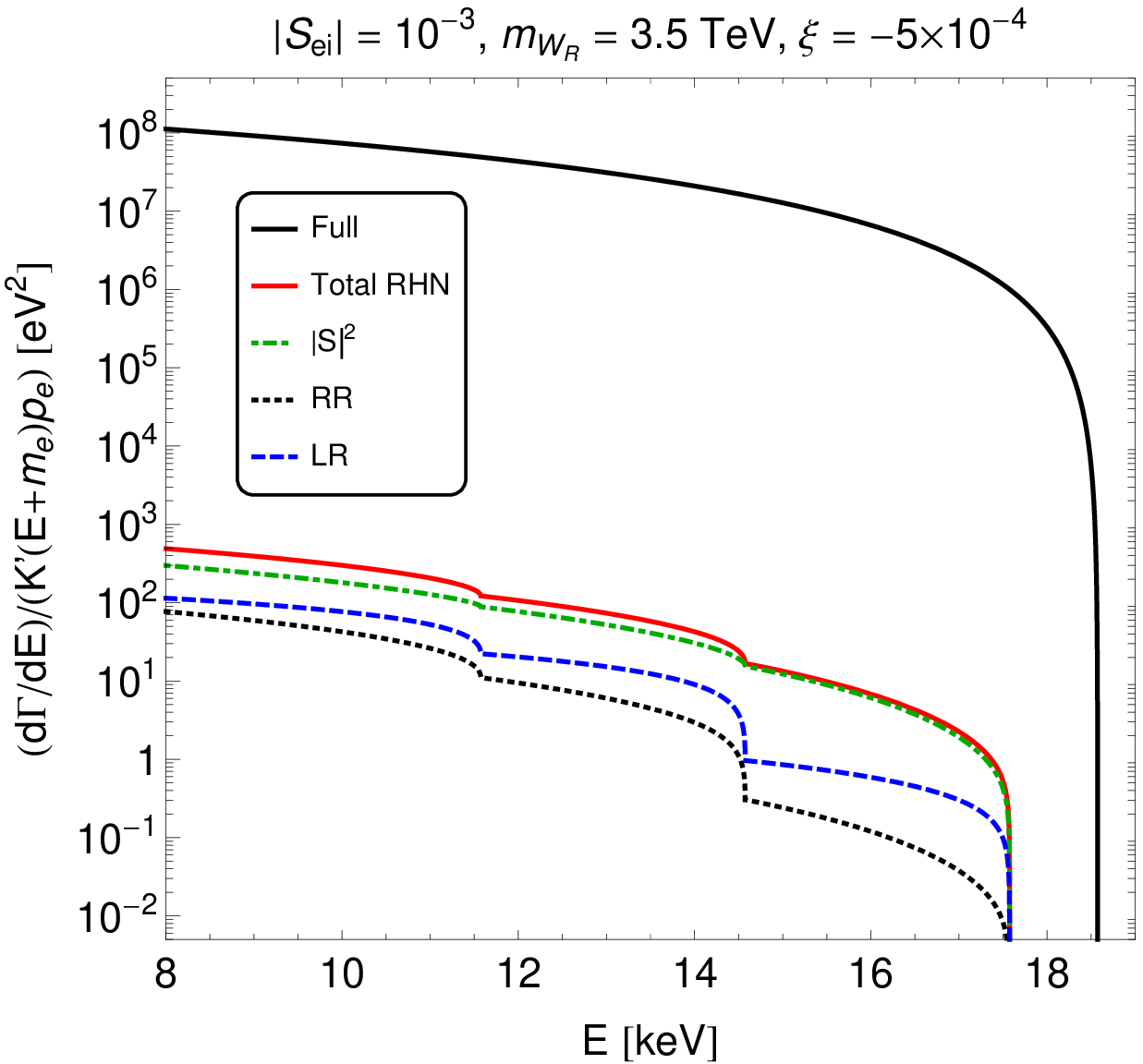} 
 \includegraphics[width=0.48\textwidth]{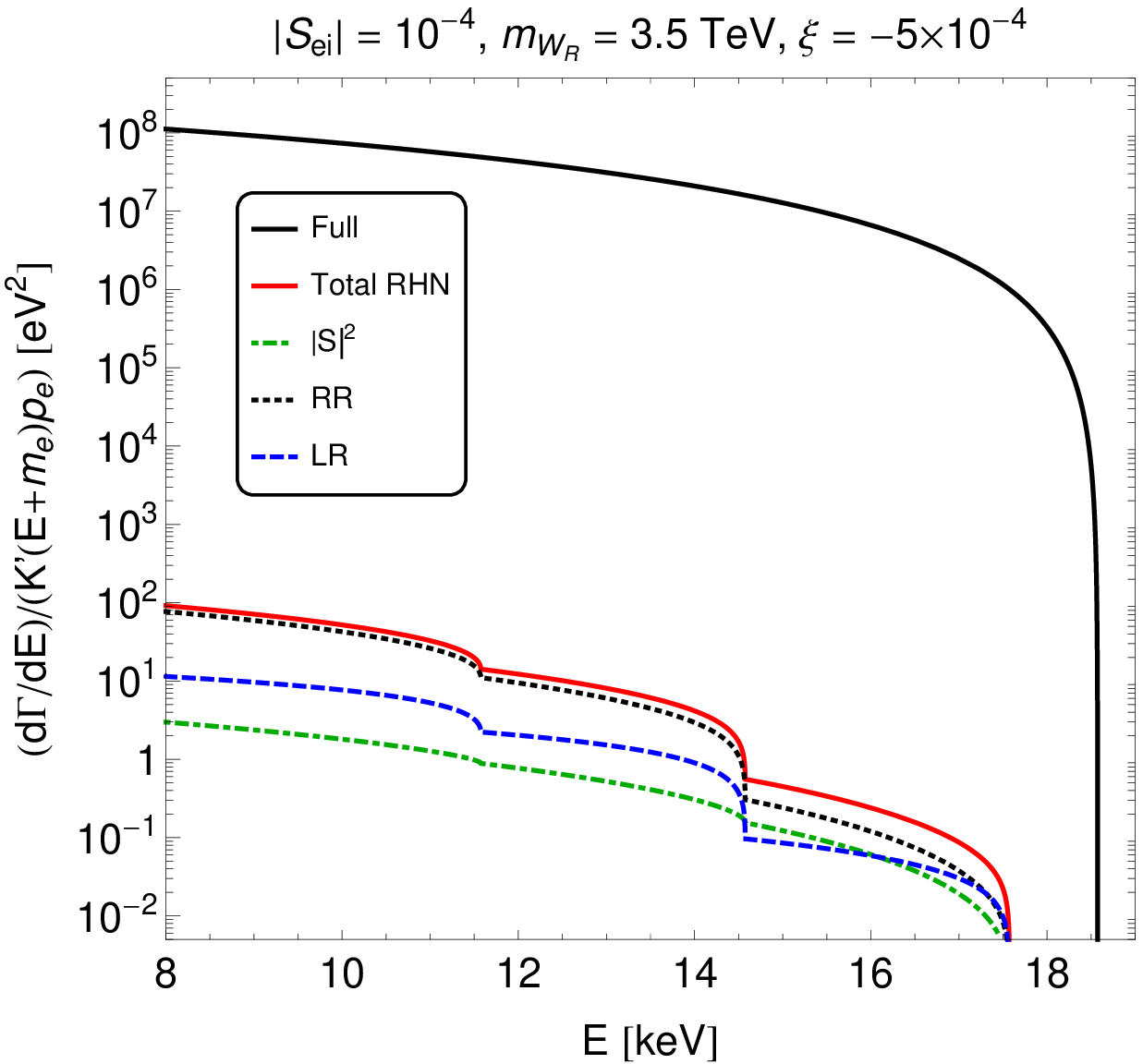}
 \caption{Differential energy spectrum $\dd \Gamma/\dd E$ of the electron in the presence of three sterile neutrinos, divided by known functions. The full spectrum including right-handed currents is shown in black; the small kinks from right-handed neutrinos (red) are the sum of the active--sterile contribution $\propto |S_{e j}|^2$ (green dot-dashed), the RR coupling from Eq.~\eqref{eq:diff_spectrum_rr} (black dotted), and the LR interference from Eq.~\eqref{eq:diff_spectrum_lr} (blue dashed). Here $|V_{ei}|=|U_{ei}|$, $\mwr = \unit[3.5]{TeV}$, $\xi=-5\times 10^{-4}$, $M_{1,2,3}=\{7,4,1\}$~keV, and $|S_{ei}|=10^{-3}$ ($|S_{ei}|= 10^{-4}$) in the left (right) plot. CP phases have been set to zero.
}
\label{fig:kurie_all}
\end{figure}

Finally, there is the contribution to beta decay purely by right-handed currents, $(\dd \Gamma/\dd E)_{RR}$. Since we are only interested in the effects of keV sterile neutrinos on the beta spectrum, we omit the RR contribution to the light neutrinos and to lowest order simply obtain
\begin{align}
\begin{split}
 \left(\frac{\dd \Gamma}{\dd E}\right)_{RR} &\simeq K'(E+m_e) p_e (E_0-E)\left[  \frac{\mwl^4}{\mwr^4}+   \tan^2 \xi + 2 C  \frac{\mwl^2}{\mwr^2} \tan\xi \cos \alpha\right]\\[1.5mm]
  &\quad  \times \sum_i |V_{ei}|^2\sqrt{(E_0-E)^2 - M_i^2}\,\Theta(E_0-E-M_i) \,,
\end{split}
\label{eq:diff_spectrum_rr}
\end{align}
stemming from the diagrams in Figs.~\ref{fig:beta_RR} and \ref{fig:beta_LR} and their interference.
Having derived the spectral contribution of keV neutrinos to second order in the small parameters $S_{e i}$, $\mwl^2/\mwr^2$, and $\xi$, we can define a modified Kurie function via
\begin{equation}
 K(E) = \sqrt{\frac{(\dd \Gamma/\dd E)_{LL}+(\dd \Gamma/\dd E)_{LR} +(\dd \Gamma/\dd E)_{RR}}{K'(E+m_e)p_e}}\, ,
 \label{eq:kuriemod}
\end{equation}
which reduces to the standard one from Eq.~\eqref{eq:standardkurie} in the absence of right-handed currents. 
The small kinks at $E = E_0 - M_i$ in the beta spectrum are illustrated in the logarithmic plot in Fig.~\ref{fig:kurie_all}; it might be possible to see the effects of keV-scale sterile neutrinos in KATRIN, as long as the sensitivity is high enough (see for instance Ref.~\cite{deVega:2011xh}).
In Fig.~\ref{fig:kurie_all} we indicate the full beta spectrum and the much smaller total contribution of the sterile neutrinos. The latter can be further divided into the standard active--sterile mixing contribution given by $|S_{e j}|^2$ in $(\dd \Gamma/\dd E)_{LL}$, the purely right-handed current mediated production given by $|V_{e j}|^2$ in $(\dd \Gamma/\dd E)_{RR}$, and the intermediate LR interference terms given by $S_{e j} V_{e j}$ from $(\dd \Gamma/\dd E)_{LR}$. Depending on the parameters, either the active--sterile or RR contribution can dominate, while the total effect on the spectrum is, of course, expected to be small.

Let us derive some analytical approximations to get a better feeling for the expected signatures.
Since $|(\dd \Gamma/\dd E)_{LR}|,(\dd \Gamma/\dd E)_{RR} \ll (\dd \Gamma/\dd E)_{LL}$, we can approximate $K(E)$ in the form
\begin{align}
 K(E) \simeq \sqrt{\frac{(\dd \Gamma/\dd E)_{LL}}{K'(E+m_e)p_e}} \left( 1 +\frac{1}{2} \frac{(\dd \Gamma/\dd E)_{LR}}{(\dd \Gamma/\dd E)_{LL}}+\frac{1}{2} \frac{(\dd \Gamma/\dd E)_{RR}}{(\dd \Gamma/\dd E)_{LL}} \right) .
\end{align}
Even more useful, however, is to expand in \emph{all} the small quantities, namely $\mwl^2/\mwr^2,|\xi|, |S_{e i}| \ll 1$, resulting in the modified Kurie function
\begin{align}
 K(E) \simeq K(E)_{\rm std} \left\{ 1 +\frac{1}{2}\sum_i G_i (E, m_i) \,\Theta(E_0-E-m_i) +\frac{1}{2} \sum_i F_i (E, M_i) \,\Theta(E_0-E-M_i)  \right\} .
 \label{eq:kurietaylor}
\end{align}
The energy-dependent coefficients $G_i (E, m_i)$ and $F_i (E, M_i)$ parametrize the shape and height of the bumps and kinks that appear on top of the standard Kurie function $K(E)_{\rm std} \simeq E_0 - E$ due to the neutrino mass eigenstates $\nu_i$ and $N_i$, and can be read off our expressions for $(\dd \Gamma/\dd E)_{LL,LR,RR}$. A similar expansion can be made for the full spectrum $\dd \Gamma/\dd E$ (in analogy to Ref.~\cite{deVega:2011xh}) instead of the Kurie function, without any qualitative differences: 
\begin{align}
 \frac{\dd \Gamma}{\dd E} \simeq \left(\frac{\dd \Gamma}{\dd E}\right)_{\rm std} \left\{1 +\sum_i G_i (E, m_i) \,\Theta(E_0-E-m_i) + \sum_i F_i (E, M_i) \,\Theta(E_0-E-M_i)\right\} .
\end{align}
We are not concerned with the spectral changes close to the beta endpoint $E_0$, seeing as those have already been studied in Ref.~\cite{Bonn:2007su}. We can therefore ignore the coefficients $G_i$, or redefine them into $(\dd \Gamma/\dd E)_{\rm std} $, without disturbing our analysis. Let us rather discuss the behavior of the spectrum in presence of keV sterile neutrinos and right-handed currents, for which we find the coefficients
\begin{subequations}
\label{eq:Fj}
\begin{align}
 F_j (E, M_j) &\simeq  \left[ |S_{e j}|^2 +|V_{e j}|^2 \left(  \frac{\mwl^4}{\mwr^4}+  \tan^2 \xi + 2 C \frac{\mwl^2}{\mwr^2} \tan\xi \cos \alpha \right) \right]\sqrt{ 1 - \frac{M_j^2}{(E_0-E)^2}}  \label{eq:FjfromLLandRR}\\
 &\quad +\frac{M_j}{E_0-E}\left[ {\rm Re} \left\{-2 S_{e j} V_{e j} \left(\frac{m_{W_L}^2}{m_{W_R}^2} + C \tan \xi e^{i\alpha}\right) \right\}  \right] \sqrt{ 1 - \frac{M_j^2}{(E_0-E)^2}} \, , \label{eq:FjfromLR}
\end{align}
\end{subequations}
to second order in  $\mwl^2/\mwr^2$, $|\xi|$, and $|S_{e i}|$, further neglecting terms of order $(\mwl^2/\mwr^2)(E/m_e)$.
The term $\propto |S_{e j}|^2$ in Eq.~\eqref{eq:FjfromLLandRR} is again just the standard active--sterile mixing contribution from $(\dd \Gamma/\dd E)_{LL}$ in Eq.~\eqref{eq:diff_spectrum_ll_heavy} (green dot-dashed line in Fig.~\ref{fig:kurie_all}), present even without right-handed currents. The terms $\propto |V_{e j}|^2$ in Eq.~\eqref{eq:FjfromLLandRR} originate from the diagrams in Figs.~\ref{fig:beta_RR} and \ref{fig:beta_LR} and their interference, calculated in $(\dd \Gamma/\dd E)_{RR}$ in Eq.~\eqref{eq:diff_spectrum_rr} and illustrated as black dotted lines in Fig.~\ref{fig:kurie_all}.
Finally, the LR interference terms $\propto S_{e j} V_{e j}$ in Eq.~\eqref{eq:FjfromLR} arise by interference of the standard diagram (Fig.~\ref{fig:beta_LL}) with the right-handed current diagrams in Figs.~\ref{fig:beta_LR} and~\ref{fig:beta_RR}, calculated in Eq.~\eqref{eq:diff_spectrum_lr} [$(\dd \Gamma/\dd E)_{LR}$]. Those require a neutrino mass-flip and therefore have a different energy-dependence than the LL and RR terms.
In the next section we will discuss the effects of this modified beta spectrum and compare it to existing constraints.

\subsection{Shapes and expected limits}
\label{sec:expected_limits}

In the previous section we derived the beta spectrum in the presence of keV sterile neutrinos and right-handed currents.
As expected, the standard active--sterile contribution $\propto |S_{e j}|^2$ is now accompanied by right-handed current mediated terms such as $|V_{e j}|^2 \mwl^4/\mwr^4$, as well as left--right mixing interference terms $\propto V_{e j} S_{e j}  \mwl^2/\mwr^2 $, which can be of similar magnitude. 
An interesting feature of the right-handed current contribution is the different energy dependence of the two terms in Eqs.~\eqref{eq:FjfromLLandRR} and~\eqref{eq:FjfromLR}, which stems from the required neutrino mass flip.\footnote{This was already noted in Refs.~\cite{Bonn:2007su, Stephenson:2000mw} for the right-handed-current contributions to active neutrinos.} As a result, the \emph{shape} $S(E)$ on top of the standard beta spectrum is described by a \emph{two}-parameter function in the presence of right-handed currents
\begin{align}
\begin{split}
 S(E) &\equiv \frac{\dd \Gamma/\dd E -  (\dd \Gamma/\dd E)_{\rm std}  }{ (\dd \Gamma/\dd E)_{\rm std}  } \simeq  2\, \frac{K(E) -  K(E)_{\rm std} }{ K(E)_{\rm std} }\\
  &\simeq \left( a   + b \frac{M}{E_0-E} \right) \sqrt{ 1 - \frac{M^2}{(E_0-E)^2}} \, \Theta(E_0-E-M) \,,
\end{split}
 \label{eq:shapes}
\end{align}
assuming for simplicity just one keV neutrino with mass $M$.
For abbreviation purposes we defined $a$ and $b$ as the terms inside the square brackets in Eq.~\eqref{eq:FjfromLLandRR} and Eq.~\eqref{eq:FjfromLR}, respectively.
Since the $b$ term originates from interference terms (to be specific: LR interference), we always have $0 \leq |b| \leq a$.

\begin{figure}
 \centering
  \includegraphics[height = 4.5cm]{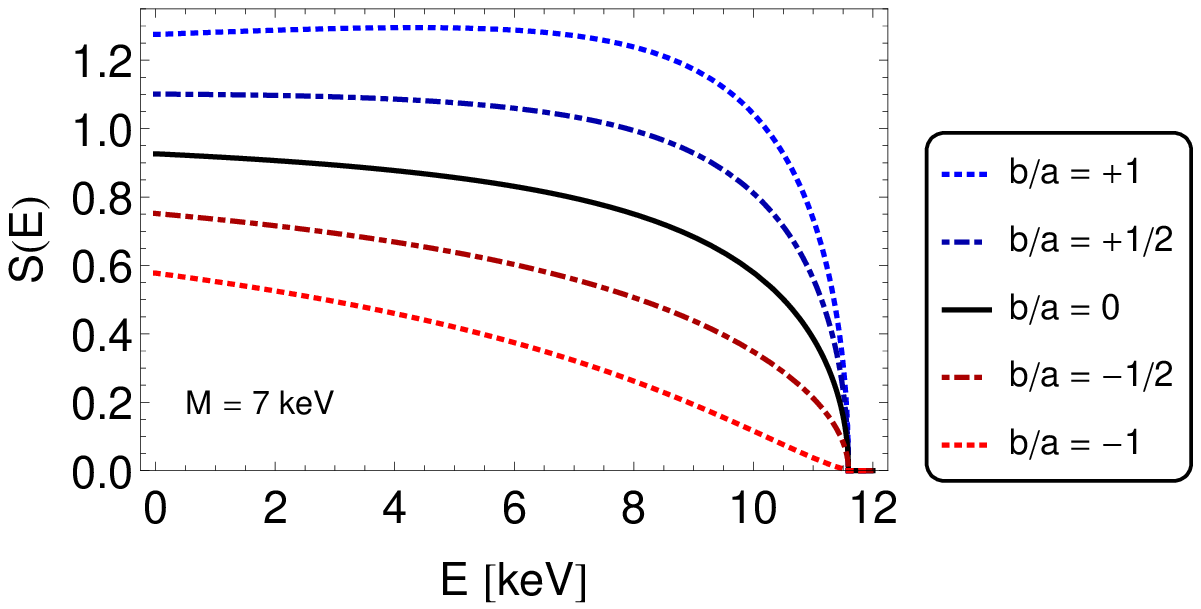}\hspace{-2ex}
  \includegraphics[height = 4.5cm]{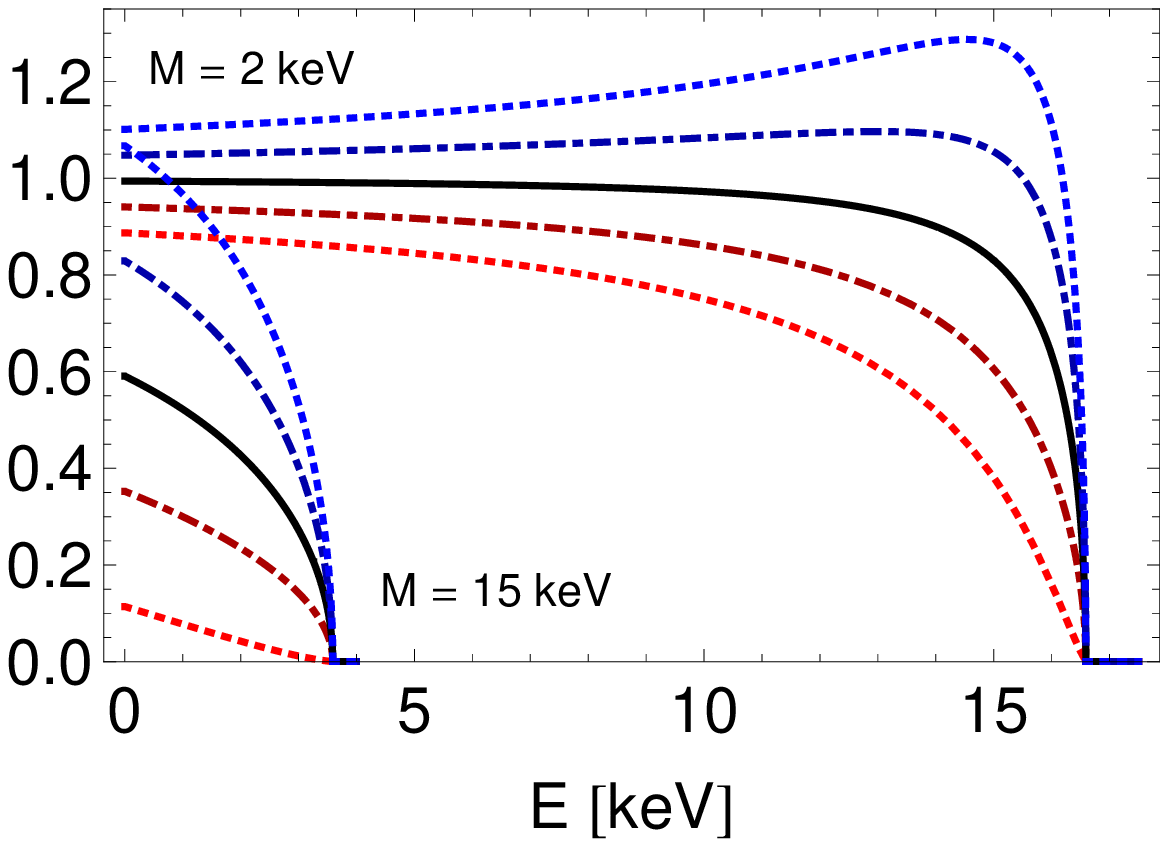}
 \caption{Modified beta spectrum $S(E) =  [\dd \Gamma/\dd E -  (\dd \Gamma/\dd E)_{\rm std}  ]/ (\dd \Gamma/\dd E)_{\rm std} $ from Eq.~\eqref{eq:shapes} due to a sterile neutrino with mass $M$. The overall magnitude is in arbitrary units, setting $a=1$, while $b/a \in [-1,+1]$ parametrizes the LR interference, be it constructive ($b>0$, blue) or destructive ($b<0$, red). The black line denotes the standard shape without right-handed currents, just active--sterile mixing.}
\label{fig:shapes}
\end{figure}

Depending on the relative size and sign of $a$ and $b$, a variety of different shapes arise, illustrated in Fig.~\ref{fig:shapes}. 
Destructive LR interference, i.e.~$b < 0$, merely flattens the spectrum. In particular, \emph{complete} destructive interference is not possible because of the additional factor $M/(E_0-E) < 1$ in Eq.~\eqref{eq:shapes}.
For $b>0$, an interesting qualitative change can occur:
Compared to the standard active--sterile contribution, the right-handed-current interference term decreases for small $E$, i.e.~large neutrino energy $E_\nu = E_0-E$, because of the required mass flip $M/E_\nu$. 
As a result, the left--right mixing contribution can lead to a \emph{maximum} in the electron's spectrum $S(E)$, a unique feature of the right-handed currents under consideration here.
The maximum appears for $b > 0$ at
\begin{align}
 \frac{E_0 - E}{M} = \frac{a+ \sqrt{a^2 + 8 b^2 }}{2 b} \,.
 \label{eq:maximum}
\end{align}
For the maximum to lie in the physical range of the spectrum, i.e.~in $0 \leq E \leq E_0$, the values must further satisfy the inequality
\begin{align}
 \frac{M}{E_0} \leq \frac{2 (b/a)}{1+ \sqrt{1 + 8 (b/a)^2 }}  = \begin{cases}
                                                           1/2 \text{ for } b = a\,,\\
                                                           b/a + \mathcal{O}(b^3/a^3) \text{ for } b \ll a\,.
                                                          \end{cases}
 \label{eq:physicalmaximum}
\end{align}
In particular, only sterile neutrinos with mass $M \leq E_0/2 \simeq \unit[9.3]{keV}$ can result in a spectral peak in the electron spectrum $S(E)$ at KATRIN, and that is under the assumption of full constructive interference of the left- and right-handed contributions ($a \simeq b$).\footnote{Including the terms of order $E/m_e$ that we dropped in our analytic approximation lowers the neutrino mass required for a maximum further to $\unit[9.1]{keV}$.}
Even then, the maximum can easily be too shallow to be observable, especially for large $M$, as can be seen in Fig.~\ref{fig:shapes} (left) for $M=\unit[7]{keV}$.
For partial interference, say $b/a \sim 0.1$, a peak is only possible for $M \lesssim \unit[1.8]{keV}$ neutrino masses. The right-handed current induced peak in the sterile neutrino kink is hence a good discovery tool only for small masses and large constructive interference. If there is no background effect/systematics that could lead to such a peak in the spectrum $S(E)$~\cite{Mertens:2013cias}, this would make it the prime discovery channel for right-handed currents.
Also note that, due to the mass flip factor $M/E_\nu$, the maximal enhancement at $b = a$ does not yield the factor 2 one could expect from constructive interference, but rather $3\sqrt{3}/4 \simeq 1.3$ at maximum (see Fig.~\ref{fig:shapes}).

Without a dedicated experimental sensitivity study in the parameter space $a$--$b$--$M$, we have to satisfy ourselves with estimates. For $b= 0$, we can immediately translate the expected limit from Eq.~\eqref{eq:KATRINsensitivity} to include the additional left--right interactions, since they share the same energy dependence: $a < \epsilon$. Nonzero $b$ introduces a different energy dependence that complicates matters; ignoring the modified \emph{shape}, we can approximately estimate the \emph{height} of the kink as $a + b M/E_0$, following the Taylor expansion
\begin{align}
a   + b \frac{M}{E_0-E} = a + b \frac{M}{E_0} + b \frac{M}{E_0} \frac{E}{E_0} + \dots 
\end{align}
For the shapes which feature an actual maximum at $E= E_{\rm max} \neq 0$ [given in Eq.~\eqref{eq:maximum}], one could define the height of the kink at that point. This would increase the height by at most a factor $1.3$, and subsequently lower the expected reach by $1/1.3\simeq 0.77$ compared to our height definition at $E=0$. The true reach should be somewhere in between.
So, in the presence of right-handed currents, we expect a sensitivity to the height of the kink 
\begin{align}
 a + b M/E_0 \simeq \epsilon \,,
\label{eq:LRsensitivity}
\end{align}
potentially improved by a shape analysis. For small masses $M$ and large positive $b\simeq a$, the sensitivity might be better, as discussed above, up to $\epsilon/1.3$.
Note that the limit now depends explicitly on the sterile neutrino mass $M$.
Inserting the left--right parameters for $a$ and $b$ from Eq.~\eqref{eq:Fj} makes the expression a little messy, so we omit it. Nevertheless, Eq.~\eqref{eq:LRsensitivity} is the main result of this work; it provides an estimated reach of a future KATRIN-like experiment to right-handed currents, given that it has a sensitivity $\epsilon$ to the standard active--sterile mixing [cf.~Eq.~\eqref{eq:KATRINsensitivity}].

For small $M$, $b$ increases/decreases the height by only a factor of order 1, so we can estimate $a \lesssim \epsilon$ even for $b\neq 0$. The effect (and limit) on $a$ is then identical to the standard case of active--sterile contributions studied in Ref.~\cite{Mertens:2013cias}, so we expect possible limits from a KATRIN-like experiment of order
\begin{align}
 \theta_{\rm eff}^2 \equiv \left[ |S_{e j}|^2 +|V_{e j}|^2 \left(  \frac{\mwl^4}{\mwr^4}+ \tan^2 \xi+ 2 C  \frac{\mwl^2}{\mwr^2} \tan \xi \cos \alpha \right) \right] \sim \epsilon 
 \label{eq:sensitivity}
\end{align}
for sterile neutrino masses in the range $\unit[1]{keV}\lesssim M_j \lesssim \unit[17]{keV}$, depending on the experimental setup. Here we defined the parameter $\theta_{\rm eff}^2$ ($a$ in the above analysis), which approximately describes the height of the sterile-neutrino kink and replaces the standard active--sterile neutrino mixing angle in the presence of right-handed currents. 
Having neglected the LR interference terms, i.e.~$b$, the above limit should be fine for the cases of LL dominance ($|S_{e j}| \gg |V_{e j}|\mwl^2/\mwr^2$), RR dominance ($|S_{e j}| \ll |V_{e j}|\mwl^2/\mwr^2$), for small interference due to the phases
\begin{align}
2 \left| {\rm Re} \left[ \left(\frac{m_{W_L}^2}{m_{W_R}^2} + C \tan \xi e^{i\alpha}\right) S_{e j} V_{e j} \right]\right| \ll
 |S_{e j}|^2 +|V_{e j}|^2 \left(  \frac{\mwl^4}{\mwr^4}+  \tan^2 \xi + 2 C \frac{\mwl^2}{\mwr^2} \tan\xi \cos \alpha \right) ,
\end{align}
and, order-of-magnitude wise, for small sterile neutrino masses, say $M\ll \unit[5]{keV}$.
The cases of large masses and strong interference require the full usage of the expected limit from Eq.~\eqref{eq:LRsensitivity}, and preferably a dedicated experimental study. We will in the following naively employ the expected sensitivity on $\theta_{\rm eff}^2$ from Eq.~\eqref{eq:sensitivity}, as this allows for an easier comparison to existing constraints.

\section{Existing limits}
\label{sec:existing_limits}

\begin{figure}
 \centering
  \includegraphics[scale = 0.7]{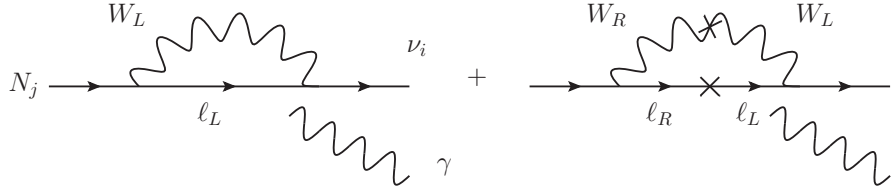}
 \caption{Neutrino decay $N_j \to \nu_i \gamma$ via active--sterile mixing (left) and left--right mixing (right). The photon line couples to any of the charged particles ($\ell$ or $W$), the crosses denote either $W_L$--$W_R$ mixing (proportional to $\xi$) or a charged-lepton mass insertion (proportional to $m_\ell$).}
\label{fig:neutrino_decay_LR}
\end{figure}

Let us collect some constraints on the parameters in $\theta_{\rm eff}^2$ [Eq.~\eqref{eq:sensitivity}] in order to evaluate the potential impact of KATRIN's limits, as derived in the last section. We first focus on the case \emph{without} right-handed currents: for a keV sterile neutrino $N_j$ to be stable enough to be dark matter in the first place, the mixing angle has to be sufficiently small to suppress the decay $N_j \to 3\nu$. An even stronger bound can be obtained by considering the loop-induced decay  $N_j \to \nu \gamma$ [Fig.~\ref{fig:neutrino_decay_LR} (left)], as it would produce a mono-energetic photon with energy $E_\gamma \simeq M_j/2$ in the X-ray range.
The non-observation of such a spectral X-ray peak from astrophysical objects with large dark matter density then puts limits on this decay rate. A simplified bound sufficient for our purposes reads~\cite{Boyarsky:2009ix} 
\begin{align}
 \Gamma (N_j \to \nu \gamma)\simeq \frac{9 G_F^2 \alpha_{\rm em}}{256 \pi^4}  \sum_\alpha |S_{\alpha j}|^2 M_j^5 \lesssim 10^{-26}\,\unit{s^{-1}} \,,
\end{align}
a more detailed overview of limits can be found in Ref.~\cite{Drewes:2013gca}. This translates to the parameters of interest as 
\begin{align}
 \theta^2_j \equiv \sum_\alpha |S_{\alpha j}|^2 \lesssim 1.8 \times 10^{-5} \left( \frac{\unit[1]{keV}}{M_j} \right)^5 .
 \label{eq:kevdecay}
\end{align}
This holds in particular for the electron component relevant for $\theta_{\rm eff}^2$ due to $|S_{e j}|^2 \leq \theta^2_j$. Without right-handed currents, the effective mixing angle $\theta_{\rm eff}^2$ relevant for KATRIN is thus always smaller than the effective mixing angle $\theta_j^2$ relevant for $N_j \to \nu\gamma$ decay: $\theta_{\rm eff}^2 \leq \theta_j^2$. 
Depending on the DM production mechanism, additional lower bounds on the neutrino mass can be obtained from phase space arguments or structure formation; the most important (and robust) for our case being the Tremaine--Gunn~\cite{Tremaine:1979we} bound $M_j \gtrsim \unit[1]{keV}$~\cite{Boyarsky:2009ix}.
With the X-ray bound from Eq.~\eqref{eq:kevdecay} in mind, a sensitivity $\epsilon$ of $10^{-8}$ ($10^{-6}$) to $\theta_{\rm eff}^2 = |S_{e j}|^2$ in KATRIN can only probe sterile neutrino masses below $\unit[4.5]{keV}$ ($\unit[1.8]{keV}$). 
Recent tantalizing hints~\cite{Bulbul:2014sua,Boyarsky:2014jta} for a dark matter particle with mass $M\simeq \unit[7]{keV}$ and mixing angle $\theta_{\rm eff}^2 \leq \theta^2 \sim 10^{-10}$--$10^{-11}$ are therefore unfortunately beyond KATRIN's expected sensitivity. As we will see in the following, the inclusion of right-handed currents can nevertheless lead to detectable signals in KATRIN, even in the case of such small active--sterile mixing angles.

Turning on the right-handed currents, we find a stringent constraint from supernova cooling for light ($M_j \lesssim \unit[1]{MeV}$) right-handed neutrinos~\cite{Barbieri:1988av,Nemevsek:2012cd}, which equates to
\begin{align}
 |V_{e j}|^2 \left(  \frac{\mwl^4}{\mwr^4} + \xi^2\right) \lesssim 10^{-10} \,.
 \label{eq:supernova}
\end{align}
Such tiny values would render the direct RR contribution invisible in KATRIN and also severely suppress the LR interference terms. (A corresponding supernova bound on $S_{e j}$ is far more difficult to establish, only bounds on $S_{\tau j}$ have been obtained in Ref.~\cite{Raffelt:2011nc}.) Let us nevertheless carry on with our analysis, looking for more robust constraints on the relevant left--right parameters.
The additional interactions can lead to an increased rate for $N_j\to \nu \gamma$, subject to the X-ray constraints mentioned above. For $\xi = 0$, the right-handed gauge bosons $W_R$ generate the additional radiative decay channel $N_j \to \overline{\nu} \gamma$, with tiny branching ratio
\begin{align}
\frac{\Gamma (N_j \to \overline{\nu} \gamma)}{\Gamma (N_j \to \nu \gamma)} \simeq \frac{\mwl^4}{\mwr^4} \frac{(V^\dagger T T^\dagger V)_{j j}}{(S^\dagger U U^\dagger S)_{j j}}\simeq \frac{\mwl^4}{\mwr^4} \,,
\end{align}
using $T\simeq - V S^\dagger V_\nu$ from Eq.~\eqref{eq:mix_matrices}. The RR contribution to neutrino decay is hence automatically sufficiently suppressed if Eq.~\eqref{eq:kevdecay} holds, independent of e.g.~$V_{e j}$. For $\xi\neq 0$ on the other hand, left--right interference can significantly enhance the decay rate, because the required mass flip on the charged-fermion line contributes a factor $m_\ell^2/M_j^2 \gg 1$~\cite{Bezrukov:2009th} (see Fig.~\ref{fig:neutrino_decay_LR}):
\begin{align}
 \Gamma (N_j \to \nu \gamma ) \simeq \frac{9 G_F^2 \alpha_{\rm em}}{256 \pi^4}  M_j^3\sum_{\ell = e, \mu, \tau} \left| S_{\ell j} M_j - \frac{8}{3}\, \xi e^{i\alpha} \,V_{\ell j}^* m_\ell \right|^2 .
\end{align}
Barring cancellations, this leads to stringent bounds on the $W$ boson mixing angle:
\begin{align}
 \xi^2 \lesssim 9\times 10^{-19} \frac{m_\tau^2}{\sum_\ell m_\ell^2 |V_{\ell j}|^2} \left(\frac{\unit[1]{keV}}{M_j}\right)^3 .
 \label{eq:xibound}
\end{align}
Due to the (approximate) unitarity of $V$, $\sum_\ell |V_{\ell j}|^2 \simeq 1$, the bound is weakest for the flavor structure $V_{e j} = 1$, $V_{\tau j} = V_{\mu j} = 0$, with $\xi^2 \lesssim 10^{-11} (\unit{keV}/M_j)^3$. Even then, the $\xi^2$ contribution to KATRIN's $\theta_{\rm eff}^2$ is far below the expected sensitivity. 
Also note that the combination of parameters relevant for beta decay, in particular the LR interference terms, can be found in the electric dipole moment of the electron, which depends at one loop on the parameters~\cite{Nieves:1986uk}
\begin{align}
 d_e \simeq 4 \times 10^{-32} \unit[e]{cm} \left( \frac{\xi}{10^{-5}} \right) \sum_{j=1}^3 \left(\frac{M_j}{\unit[1]{keV}} \right) G\left( \frac{M_j^2}{\mwl^2}\right) {\rm Im} \left( S_{e j} V_{e j} e^{i \alpha}  \right) ,
\end{align}
with a loop function $1/2 \leq G(x)< 2$. The contribution of a keV sterile neutrino is suppressed far below the current limit of $|d_e| < \unit[10^{-27} e]{cm}$~\cite{Beringer:1900zz}, so this constraint is not \emph{directly} relevant for KATRIN, seeing as it furthermore depends strongly on CP phases. 

Collider searches for the missing energy of a light sterile neutrino $N_j$ following the decay $W_R\to e N_j$ can be sensitive to the matrix element $V_{e j}$ we are interested in. If all three right-handed neutrinos $N_j$ are much lighter than the $W_R$, unitarity results in the $95\%$~C.L.~limit $\mwr \gtrsim \unit[2.5]{TeV}$~\cite{Chatrchyan:2012meb}, independent of the entries $V_{\alpha j}$. In this case, the product relevant for KATRIN
\begin{align}
  \theta_{\rm eff}^2 \, \supset \, |V_{e j}|^2 \left(\frac{\mwl}{\mwr}\right)^4 \simeq  1.1\times 10^{-6} \, |V_{e j}|^2 \left(\frac{\unit[2.5]{TeV}}{\mwr}\right)^4
\end{align}
is within the expected sensitivity $\epsilon$.\footnote{Preliminary results from the full $\sqrt{s} = \unit[8]{TeV}$ LHC run indicate improved limits around $\mwr \gtrsim \unit[3.3]{TeV}$~\cite{atlasnote,CMS:2013rca}. This corresponds to $\mwl^4/\mwr^4 \simeq 3.5\times 10^{-7}$, still within KATRIN's expected reach.}
In case of a spectrum $M_1 \ll \mwr < M_{2,3}$, the branching ratio $W_R \to e N_1$ is no longer $\simeq 1/12$, but depends on the entry of interest in the form $\simeq |V_{e 1}|^2/10$. A small $V_{e 1}$ then allows for a smaller $\mwr$, e.g.~$\mwr \gtrsim \unit[1.8]{TeV}$ for $|V_{e 1}|\simeq 0.28$. In this case, the product relevant for KATRIN is still within reach, $|V_{e 1}|^2 \mwl^4/\mwr^4\simeq 3\times 10^{-7}$.
Note that these collider limits are derived under the assumption that the right-handed quark mixing matrix is identical to the left-handed one (up to phases), in particular with large $u d$ entry $|V^R_{u d}| = |V^L_{u d}| \simeq 1$. This assumption can be relaxed in more general left--right models without an additional discrete charge conjugation or parity symmetry~\cite{Langacker:1989xa}, and would loosen the above constraints.
Overall, we therefore expect KATRIN to provide limits on keV sterile neutrinos with right-handed currents competitive with current LHC limits. In particular, KATRIN can probe the purely leptonic right-handed currents, without any assumptions on the right-handed quark mixing matrix.

This concludes our survey of existing constraints on the parameters relevant for KATRIN's keV sterile neutrino search.
If taken seriously, the astrophysical X-ray and supernova constraints from Eqs.~\eqref{eq:xibound} and~\eqref{eq:supernova} would render the left--right contribution to KATRIN's $\theta_{\rm eff}^2$ parameter unobservably small, and consequently restrict the parameter space to be probed in KATRIN to large active--sterile mixing $|S_{e j}| \gtrsim \sqrt{\epsilon}\simeq  10^{-4}$--$10^{-3}$ [corresponding to small masses $M_j \lesssim \unit[4.5]{keV}$ with Eq.~\eqref{eq:kevdecay}].
Since the supernova cooling bound from Eq.~\eqref{eq:supernova} arguably suffers from unknown systematic effects, we will ignore it in the following; in any case, it would be desirable to probe the $V_{e j}$ coupling in a terrestrial, well-understood environment, which is precisely where KATRIN comes in, as it can probe the relevant parameter $|V_{e j}|^2 \mwl^4/\mwr^4$ to $\epsilon \simeq 10^{-6}$--$10^{-8}$ (compared to $10^{-10}$ from supernovae and $\sim 10^{-6}$ from LHC~\cite{Chatrchyan:2012meb}). Without the supernova bound, there is still the more robust X-ray limit on $\xi$ from Eq.~\eqref{eq:xibound}, which is to be taken seriously. We will consequently neglect $\xi$ compared to $\mwl^2/\mwr^2$, further simplifying our expression for $\theta_{\rm eff}^2$ from Eq.~\eqref{eq:sensitivity}.

Loosening the supernova cooling bound from Eq.~\eqref{eq:supernova} then allows for kinks over the \emph{entire} mass region sensitive in KATRIN, say $\unit[1]{keV} \lesssim M_j \lesssim \unit[17]{keV}$, dominated for the most part by RR contributions
\begin{align}
 \theta^2_{\rm eff} \simeq  |S_{e j}|^2  + 1.1\times 10^{-6} \, |V_{e j}|^2 \left(\frac{\unit[2.5]{TeV}}{\mwr}\right)^4 .
\end{align}
We stress again that the RR contributions do not have any significant impact on the sterile neutrino decay rate probed by X-ray searches, and is hence not subject to those constraints. They are independently constrained by LHC searches for $W_R$, though. In effect, the right-handed currents allow for KATRIN's effective angle $\theta_{\rm eff}^2$ to be \emph{larger} than the effective active--sterile mixing $\theta_j^2$ relevant for X-rays, contrary to the case without right-handed currents.
This makes beta decay interesting even if a warm dark matter signal with seemingly too small mixing angle were to be established by an X-ray line, the latter being insensitive to right-handed currents.
Beta decay can therefore provide important limits using well-understood laboratory experiments. In the next section we will discuss how these limits compare to existing and future bounds from neutrinoless double beta decay experiments.

\section{Comparison to neutrinoless double beta decay}
\label{sec:doublebeta}

It is interesting to compare the signatures of right-handed currents in beta decay with their analogous effects in neutrinoless double beta decay ($\obb$), which were recently studied in
Refs.~\cite{Barry:2013xxa,Huang:2013kma}. There are various possible ways that the process can occur, which can be seen by simply connecting the neutrino lines of two single beta decays from
Figs.~\ref{fig:beta_LL} and \ref{fig:beta_LRSM}. The ``standard'' case in which light active Majorana neutrinos are exchanged has an amplitude given by
\begin{equation}
 {\cal A}_\nu \simeq G_F^2 \sum_i \frac{U_{ei}^2m_i}{q^2} \equiv G_F^2 \frac{\mee}{q^2},
\end{equation}
where $|\mee|=|\sum_iU_{ei}^2m_i|$ is the effective mass [compare with Eq.~\eqref{eq:mbeta_def}] and $|q| \simeq 100$~MeV is the typical scale of momentum transfer in the process. In the LRSM with
right-handed neutrinos the structure of the amplitude is determined by the mass of the right-handed states relative to $|q|$, since the neutrino propagator behaves differently for $M_i \ll |q|$ and
$M_i \gg |q|$. The different amplitudes were discussed in Ref.~\cite{Barry:2013xxa}, where the focus was on the case of TeV-scale right-handed neutrinos with large mixing. The authors of
Ref.~\cite{Huang:2013kma} also examine the case where some of the right-handed neutrinos are above and others below the scale $|q|$, which leads to interesting phenomenology. It is precisely this case
that is relevant to our study of right-handed currents in KATRIN, since only neutrino states with mass below the $Q$-value of Tritium can be produced in the experiment. 

The amplitudes for $\obb$ can be divided in terms of the chirality of the final state electrons. We give the most general expressions and then focus on a specific right-handed neutrino mass spectrum; in each case ``light'' (``heavy'') refers to right-handed neutrinos with $M_i \ll |q|$ ($M_i \gg |q|$). With both electrons left-handed, the leading terms of the amplitude (neglecting terms with gauge
boson mixing) are
\begin{equation}
 {\cal A}_{LL} \simeq G_F^2\left[ \left(\sum_{i=1}^3U_{ei}^2m_i+\sum_i^{\rm light}S_{ei}^2M_i\right)\frac{1}{q^2}-\sum_{i}^{\rm heavy}\frac{S_{ei}^2}{M_i}\right], \label{eq:amp_LL}
\end{equation}
whereas with both electrons right-handed the amplitude is
\begin{equation}
 {\cal A}_{RR} \simeq G_F^2 \left(\frac{\mwl}{\mwr}\right)^4 \left[\left(\sum_{i=1}^3{T_{ei}^*}^2m_i + \sum_i^{\rm light}{V_{ei}^*}^2M_i\right)\frac{1}{q^2}-\sum_{i}^{\rm
heavy}\frac{{V^*_{ei}}^2}{M_i}\right]. \label{eq:amp_RR}
\end{equation}
For illustration purposes we suppress the nuclear matrix elements that accompany each term; these depend on the neutrino mass and helicity of the final states.
Mixing between light and heavy neutrinos as well as gauge boson mixing leads to diagrams with mixed helicity final states, their amplitude is
\begin{equation}
 {\cal A}_{LR} \simeq G_F^2 \left(\frac{\mwl^2}{\mwr^2}+\tan\xi\right) \left[\left(\sum_{i=1}^3U^{}_{ei} T^*_{ei}+\sum_i^{\rm light}S^{}_{ei}V^*_{ei}\right)\frac{1}{q}-\sum_i^{\rm
heavy}S^{}_{ei}V_{ei}^*\frac{q}{M_i^2}\right] ,
 \label{eq:amp_LR}
\end{equation}
where the prefactors $(\mwl^2/\mwr^2)$ and $\tan\xi$ correspond to the so-called $\lambda$- and $\eta$-diagrams, respectively. The expressions given in Eqs.~\eqref{eq:amp_LL}, \eqref{eq:amp_RR} and
\eqref{eq:amp_LR} can be further simplified depending on the specific right-handed neutrino mass spectrum.

If one considers a scenario in which one right-handed neutrino ($N_{R1}$) has a mass ($M_1$) of several keV whereas the other two are situated at the GeV scale or higher, the amplitudes become
\begin{align}
\begin{split}
 {\cal A}_{LL}|_{M_1 \ll |q|} &\simeq G_F^2\left[\left([M_L]_{ee} - \sum_{i=2}^3S_{ei}^2M_i\right)\frac{1}{q^2}-\sum_{i=2}^3\frac{S_{ei}^2}{M_i}\right] \\
 & \approx G_F^2\left(\frac{v_L}{v_R}[M_R]_{ee} - \sum_{i=2}^3S_{ei}^2M_i\right)\frac{1}{q^2}\, ,\end{split}\\[1.5mm]
 \begin{split}
 {\cal A}_{RR}|_{M_1 \ll |q|} &\simeq G_F^2 \left(\frac{\mwl}{\mwr}\right)^4 \left[\left(\sum_{i=1}^3{T_{ei}^*}^2m_i+{V_{e1}^*}^2M_1\right)\frac{1}{q^2}-\sum_{i=2}^3\frac{{V^*_{ei}}^2}{M_i}\right] \\
 & \approx G_F^2 \left(\frac{\mwl}{\mwr}\right)^4 \left({V_{e1}^*}^2M_1\frac{1}{q^2}-\sum_{i=2}^3\frac{{V^*_{ei}}^2}{M_i}\right), \end{split} \\[1.5mm]
 \begin{split}
 {\cal A}_{LR}|_{M_1 \ll |q|} &\simeq G_F^2 \left(\frac{\mwl^2}{\mwr^2}+\tan\xi\right) \left[-\sum_{i=2}^3\left(S^{}_{ei}V^*_{ei}\frac{1}{q}+S^{}_{ei}V_{ei}^*\frac{q}{M_i^2}\right)\right] \\
 &\approx G_F^2 \left(\frac{\mwl^2}{\mwr^2}+\tan\xi\right) \left(-\sum_{i=2}^3S^{}_{ei}V^*_{ei}\frac{1}{q}\right),
  \end{split}
\end{align}
where we have used Eqs.~\eqref{eq:md_alt} and \eqref{eq:md_ml_symm_relations} as well as the fact that (barring cancellations) we expect the terms proportional to $|S|^2/M_i$, $|T|^2m_i$ and
$|S|/M_i^2$ to be
suppressed. The final halflife, putting the nuclear matrix elements back in, will be
\begin{equation}
 [T_{1/2}^{0\nu}]^{-1} = G_{01}^{0\nu}\left\{|{\cal M}_\nu^{0\nu}|^2|\eta_{LL}|^2 + |{\cal M}_\nu^{0\nu}|^2|\eta^{\rm light}_{RR}|^2 + |{\cal M}_N^{0\nu}|^2|\eta^{\rm heavy}_{RR}|^2+|{\cal
M}_\lambda^{0\nu}\eta_\lambda+{\cal M}_\eta^{0\nu}\eta_\eta|^2\right\},
\end{equation}
where the phase space factor $G_{01}^{0\nu}$ and matrix elements ${\cal M}_k^{0\nu}$ ($k=\nu,N,\lambda,\eta$) can be found in Table~2 of Ref.~\cite{Barry:2013xxa}, and the dimensionless LNV
parameters with their corresponding limits are given by
\begin{align}
 |\eta_{LL}| &= \frac{1}{m_e}\left|\frac{v_L}{v_R}[M_R]_{ee} - \sum_{i=2}^3S_{ei}^2M_i\right| \lesssim 7.1 \times 10^{-7}, \\[1mm]
 |\eta^{\rm light}_{RR}| &= \frac{1}{m_e}\left(\frac{\mwl}{\mwr}\right)^4 \left|V_{e1}\right|^2 M_1 \lesssim 7.1 \times 10^{-7}\, , \label{eq:lightRR}\\[1mm]
 |\eta^{\rm heavy}_{RR}| &= m_p\left(\frac{\mwl}{\mwr}\right)^4\left|-\sum_{i=2}^3\frac{{V^*_{ei}}^2}{M_i}\right| \lesssim 7.0 \times 10^{-9}\, ,\\[1mm]
 |\eta_\lambda| &= \frac{\mwl^2}{\mwr^2}\left|-\sum_{i=2}^3S^{}_{ei}V^*_{ei}\right| \lesssim 5.7 \times 10^{-7}\, , \\[1mm]
 |\eta_\eta| &= \tan\xi\left|-\sum_{i=2}^3S^{}_{ei}V^*_{ei}\right| \lesssim 3.0 \times 10^{-9}\, . 
\end{align}
The current limits on each parameter have been calculated as in Ref.~\cite{Barry:2013xxa}, using the half-life limit from the KamLAND-Zen experiment and neglecting interference terms.

Following the arguments in Ref.~\cite{Huang:2013kma}, it turns out that only the $RR$ and $LR$ contributions are important here, since usually $v_L \ll v_R$ and the mixing $S$ is small so that the
$LL$ contribution is suppressed. If we assume that only one contribution dominates at a time we can get limits on the mass and mixing parameters of the right-handed neutrinos and thus show the
correlation between half-lives in $\obb$ and the differential electron spectrum in KATRIN, since both depend on the same parameters.

With the current limits on the left--right contributions to $\obb$ at our disposal, we can perform a comparison with the expected sensitivity of a future KATRIN-like beta-decay experiment to a keV neutrino $N_j$, as given in Eq.~\eqref{eq:sensitivity}, for convenience repeated here
\begin{align}
 \theta_{\rm eff}^2  = \left[ |S_{e j}|^2 +|V_{e j}|^2 \left(  \frac{\mwl^4}{\mwr^4}+ \tan^2 \xi + 2 C  \frac{\mwl^2}{\mwr^2} \tan \xi \cos \alpha \right) \right] \lesssim \epsilon\,.
\end{align}
Here $\epsilon = 10^{-6}$--$10^{-8}$ is subject to an $\mathcal{O}(1)$ fudge factor rooted in ruthless approximations (see discussion in Sec.~\ref{sec:expected_limits}).
The most interesting quantity from $\obb$ is $\eta_{RR}^{\rm light}$ from Eq.~\eqref{eq:lightRR}, as it depends on the same parameters as the dominant RR contribution in KATRIN
\begin{align}
 \left|V_{e j}\right|^2 \left(\frac{\mwl}{\mwr}\right)^4 = \frac{m_e}{ M_j}\,  |\eta^{\rm light}_{RR}| \simeq 3.6\times 10^{-4} \left( \frac{\unit[1]{keV}}{M_j}\right) \left(\frac{|\eta^{\rm light}_{RR}|}{7.1 \times 10^{-7}} \right)  ,
\end{align}
and is well within the expected sensitivity of $10^{-6}$--$10^{-8}$. Future $\obb$ experiments such as GERDA phase-II might push the bound on $\eta^{\rm light}_{RR}$ by another order of magnitude, but KATRIN (and the LHC) can easily provide better limits on this quantity. The other limits from $\obb$ depend on the KATRIN parameters only indirectly, e.g.~via unitarity or seesaw formulae, which makes a direct comparison of these two approaches more involved. In particular, it is unavoidably necessary to specify the parameters for the two heavier right-handed neutrinos. We leave a detailed parameter scan for future work. Nevertheless, one can expect that the additional 
interactions in the form of right-handed currents allow for a more testable signal 
of keV-ish warm dark matter neutrinos in double beta decay than without them. Recall that the 
decisive quantity for the process in case of the standard light 
neutrino exchange is $U_{ei}^2  m_i$, currently limited to 
about 0.3\,eV~\cite{inder}. 
Without right-handed currents, 
the contribution of, say, a $M=7$\,keV neutrino to the process corresponds 
to an effective mass of $\theta^2 M \simeq 70\,\mu$eV, where 
$\theta^2 \simeq 10^{-10}$ is the mixing parameter extracted 
from the recent hint for an X-ray line~\cite{Bulbul:2014sua,Boyarsky:2014jta}. 
Including right-handed currents, 
and recalling that the X-ray constraints are essentially left 
untouched by them, gives a contribution corresponding to 
$(\mwl/\mwr)^4 \, |V_{ei}|^2 M$, which can be up to 
$0.008$\,eV, a value not out of reach for future experiments~\cite{Rodejohann:2011mu}.

\begin{figure}
 \centering
  \includegraphics[scale = 0.7]{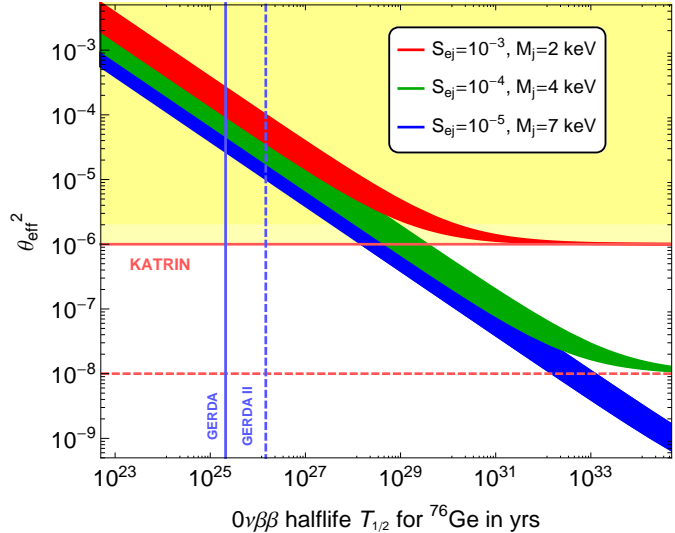}
 \caption{Effective beta-decay mixing angle $\theta^2_{\rm eff}$ as a function of $\obb$ lifetime in $^{76}$Ge for some sterile neutrino masses $M_j$ and active--sterile mixing angles $S_{e j}$ in the presence of right-handed currents, following Eq.~\eqref{eq:thetavslifetime}. The spread comes from uncertainties in the nuclear matrix elements.
 The horizontal lines denote the expected reach of a future KATRIN-like experiment, the vertical lines the lifetime limit from GERDA~\cite{Agostini:2013mzu} and expected phase-II improvement.
The shaded yellow area $\theta_{\rm eff}^2 - |S_{e j}|^2 \gtrsim 10^{-6}$ is excluded by the LHC bound $\mwr \gtrsim \unit[2.5]{TeV}$~\cite{Chatrchyan:2012meb}.}
\label{fig:KATRINvsGERDA}
\end{figure}

To show one example of a possible correlation, let us assume that $\obb$ is dominated by $\eta_{RR}^{\rm light}$, i.e.~by the keV sterile neutrino $N_j$. Neglecting further the W boson mixing $\xi$ in accordance with the X-ray bound from Eq.~\eqref{eq:xibound}, we find a direct relation between the effective beta-decay mixing angle $\theta_{\rm eff}$ and the $\obb$ lifetime, purely due to right-handed currents:
\begin{align}
 \theta_{\rm eff}^2 = |S_{e j}|^2 + \frac{m_e}{M_j} |\mathcal{M}_\nu^{0\nu}|^{-1} \left(G_{01}^{0\nu}\right)^{-\frac12} \left(T_{1/2}^{0\nu}\right)^{-\frac12} .
 \label{eq:thetavslifetime}
\end{align}
We show this expression in Fig.~\ref{fig:KATRINvsGERDA} for the germanium isotope $^{76}$Ge for a variety of mixing angles and masses. Here we indicate a $\theta_{\rm eff}^2$ sensitivity of $10^{-6}$ (horizontal line) or $10^{-8}$ (dashed horizontal line), for simplicity independent of the sterile neutrino mass. (A virtually identical plot can be made with regards to the $^{136}$Xe lifetime.) 
For the values chosen in Fig.~\ref{fig:KATRINvsGERDA}, the area above $\theta_{\rm eff}^2 \gtrsim (1$--$2)\times 10^{-6}$ is already excluded by the LHC limit $\mwr \gtrsim \unit[2.5]{TeV}$~\cite{Chatrchyan:2012meb}.\footnote{We note again that preliminary results indicate improved limits around $\mwr \gtrsim \unit[3.3]{TeV}$~\cite{atlasnote,CMS:2013rca}.}
We see that a future KATRIN-like experiment can provide important constraints on the parameter space of keV sterile neutrinos, complementary to the LHC and beyond the reach of current $\obb$ experiments.

\section{Conclusion}
\label{sec:conclusion}

Sterile neutrinos with keV masses are a well-motivated candidate for warm dark matter and require an experimental discovery machinery very different from cold dark matter candidates. Due to their mixing with active neutrinos they are expected to manifest themselves in well-known processes such as beta decay, resulting in kinks in the electron's spectrum. In this paper we discussed the impact of right-handed currents on the beta-decay phenomenology of keV sterile neutrinos, specifically on tritium beta decay at a future KATRIN-like experiment.
The additional interactions in a left--right model can enhance the \emph{height} of the kink, and left--right interference can further modify the \emph{shape} of the kink, best visible for small sterile neutrino masses, say below $\unit[5]{keV}$. 
Since the direct contribution to the spectral kink by right-handed $W_R$ bosons is not linked to the decay rate of the sterile neutrino (as opposed to $W_L$--$W_R$ mixing), appreciable signals in the beta spectrum can arise even for small active--sterile mixing angles. This significantly increases the potential discovery reach of KATRIN to warm dark matter keV sterile neutrinos, liberating it partially from X-ray constraints.
The expected sensitivity to the dominant new-physics contribution via $W_R$ exchange is beyond upcoming $\obb$ experiments, but complementary to current LHC results, making a dedicated analysis of this scenario by our experimental colleagues worthwhile.

\csection{Acknowledgements}
~This work was supported by the Max Planck Society in the project MANITOP, JH acknowledges support by the IMPRS-PTFS.

\bibliographystyle{utcaps_mod}
\bibliography{lrbib}

\providecommand{\href}[2]{#2}\begingroup\raggedright\begin{thebibliography}{10}

\bibitem{Beringer:1900zz}
{\bf Particle Data Group}, J.~Beringer {\em et al.}, ``{\em {Review of Particle
  Physics}},''
\href{http://dx.doi.org/10.1103/PhysRevD.86.010001}{Phys. Rev. {\bf D86} (2012)
   010001}.

\bibitem{Drexlin:2013lha}
G.~Drexlin, V.~Hannen, S.~Mertens, and C.~Weinheimer, ``{\em {Current direct
  neutrino mass experiments}},'' \href{http://dx.doi.org/10.1155/2013}{Adv.High
  Energy Phys. {\bf 2013} (2013)  293986},
\href{http://arxiv.org/abs/1307.0101}{{\tt arXiv:1307.0101 [physics.ins-det]}}.

\bibitem{Otten:2008zz}
E.~Otten and C.~Weinheimer, ``{\em {Neutrino mass limit from tritium beta
  decay}},''
  \href{http://dx.doi.org/10.1088/0034-4885/71/8/086201}{Rept.Prog.Phys. {\bf
  71} (2008)  086201},
\href{http://arxiv.org/abs/0909.2104}{{\tt arXiv:0909.2104 [hep-ex]}}.

\bibitem{Angrik:2005ep}
{\bf KATRIN Collaboration}, J.~Angrik {\em et al.}, ``{\em {KATRIN design
  report 2004}},''.
\url{http://inspirehep.net/record/680949/files/FZKA7090.pdf}.

\bibitem{Mertens:2013cias}
S.~Mertens, ``{\em {Search for keV neutrinos with KATRIN}}.'' {Chalonge Meudon
  Workshop}, 2013.
\newblock \url{http://chalonge.obspm.fr/CIAS13_Mertens.pdf}.

\bibitem{Minkowski:1977sc}
P.~Minkowski, ``{\em {$\mu \to e \gamma$ at a Rate of One Out of 1-Billion Muon
  Decays?}},''
\href{http://dx.doi.org/10.1016/0370-2693(77)90435-X}{Phys. Lett. {\bf B67}
  (1977)  421}.

\bibitem{GellMann:1980vs}
M.~Gell-Mann, P.~Ramond, and R.~Slansky, ``{\em {Horizontal gauge symmetry and
  masses of neutrinos}},'' in {\em {Supergravity}}, P.~{van Nieuwenhuizen} and
  D.~Freedman, eds., p.~315.
\newblock 1979.

\bibitem{Yanagida:1979as}
T.~Yanagida, ``{\em {Horizontal Gauge Symmetry And Masses Of Neutrinos}},'' in
  {\em {Proc. Workshop on the Baryon Number of the Universe and Unified
  Theories}}, O.~Sawada and A.~Sugamoto, eds., p.~95.
\newblock 1979.

\bibitem{Mohapatra:1979ia}
R.~N. Mohapatra and G.~Senjanovi{\'c}, ``{\em {Neutrino Mass and Spontaneous
  Parity Violation}},''
\href{http://dx.doi.org/10.1103/PhysRevLett.44.912}{Phys. Rev. Lett. {\bf 44}
  (1980)  912}.

\bibitem{Fukugita:1986hr}
M.~Fukugita and T.~Yanagida, ``{\em {Baryogenesis Without Grand
  Unification}},''
\href{http://dx.doi.org/10.1016/0370-2693(86)91126-3}{Phys.Lett. {\bf B174}
  (1986)  45}.

\bibitem{Bode:2000gq}
P.~Bode, J.~P. Ostriker, and N.~Turok, ``{\em {Halo formation in warm dark
  matter models}},'' \href{http://dx.doi.org/10.1086/321541}{Astrophys.J. {\bf
  556} (2001)  93--107},
\href{http://arxiv.org/abs/astro-ph/0010389}{{\tt arXiv:astro-ph/0010389
  [astro-ph]}}.

\bibitem{Shrock:1980vy}
R.~Shrock, ``{\em New tests for, and bounds on, neutrino masses and lepton
  mixing},''
\href{http://dx.doi.org/10.1016/0370-2693(80)90235-X}{Phys.Lett. {\bf B96}
  (1980)  159}.

\bibitem{deVega:2011xh}
H.~de~Vega, O.~Moreno, E.~M. de~Guerra, M.~R. Medrano, and N.~S{\'a}nchez,
  ``{\em {Role of sterile neutrino warm dark matter in rhenium and tritium beta
  decays}},''
  \href{http://dx.doi.org/10.1016/j.nuclphysb.2012.08.019}{Nucl.Phys. {\bf
  B866} (2013)  177--195},
\href{http://arxiv.org/abs/1109.3452}{{\tt arXiv:1109.3452 [hep-ph]}}.

\bibitem{Abdurashitov:2014vqa}
D.~Abdurashitov, A.~Berlev, N.~Likhovid, A.~Lokhov, I.~Tkachev, {\em et al.},
  ``{\em {Search for an admixture of sterile neutrino in the electron spectrum
  from tritium $\beta$-decay}},''
\href{http://arxiv.org/abs/1403.2935}{{\tt arXiv:1403.2935 [nucl-ex]}}.

\bibitem{Herczeg:2001vk}
P.~Herczeg, ``{\em {Beta decay beyond the standard model}},''
\href{http://dx.doi.org/10.1016/S0146-6410(01)00149-1}{Prog.Part.Nucl.Phys.
  {\bf 46} (2001)  413--457}.

\bibitem{Severijns:2006dr}
N.~Severijns, M.~Beck, and O.~Naviliat-Cuncic, ``{\em {Tests of the standard
  electroweak model in beta decay}},''
  \href{http://dx.doi.org/10.1103/RevModPhys.78.991}{Rev.Mod.Phys. {\bf 78}
  (2006)  991--1040},
\href{http://arxiv.org/abs/nucl-ex/0605029}{{\tt arXiv:nucl-ex/0605029
  [nucl-ex]}}.

\bibitem{Monreal:2009za}
B.~Monreal and J.~A. Formaggio, ``{\em {Relativistic Cyclotron Radiation
  Detection of Tritium Decay Electrons as a New Technique for Measuring the
  Neutrino Mass}},''
  \href{http://dx.doi.org/10.1103/PhysRevD.80.051301}{Phys.Rev. {\bf D80}
  (2009)  051301},
\href{http://arxiv.org/abs/0904.2860}{{\tt arXiv:0904.2860 [nucl-ex]}}.

\bibitem{Mohapatra:1974gc}
R.~Mohapatra and J.~C. Pati, ``{\em {A Natural Left-Right Symmetry}},''
\href{http://dx.doi.org/10.1103/PhysRevD.11.2558}{Phys. Rev. {\bf D11} (1975)
  2558}.

\bibitem{Pati:1974yy}
J.~C. Pati and A.~Salam, ``{\em {Lepton Number as the Fourth Color}},''
\href{http://dx.doi.org/10.1103/PhysRevD.10.275; 10.1103/PhysRevD.11.703}{Phys.
  Rev. {\bf D10} (1974)  275--289}.

\bibitem{Senjanovic:1975rk}
G.~Senjanovi{\'c} and R.~N. Mohapatra, ``{\em {Exact Left-Right Symmetry and
  Spontaneous Violation of Parity}},''
\href{http://dx.doi.org/10.1103/PhysRevD.12.1502}{Phys. Rev. {\bf D12} (1975)
  1502}.

\bibitem{Mohapatra:1980yp}
R.~N. Mohapatra and G.~Senjanovi{\'c}, ``{\em {Neutrino Masses and Mixings in
  Gauge Models with Spontaneous Parity Violation}},''
\href{http://dx.doi.org/10.1103/PhysRevD.23.165}{Phys. Rev. {\bf D23} (1981)
  165}.

\bibitem{Appelquist:2003uu}
T.~Appelquist and R.~Shrock, ``{\em {Dynamical symmetry breaking of extended
  gauge symmetries}},''
  \href{http://dx.doi.org/10.1103/PhysRevLett.90.201801}{Phys.Rev.Lett. {\bf
  90} (2003)  201801},
\href{http://arxiv.org/abs/hep-ph/0301108}{{\tt arXiv:hep-ph/0301108
  [hep-ph]}}.

\bibitem{Bezrukov:2009th}
F.~Bezrukov, H.~Hettmansperger, and M.~Lindner, ``{\em {keV sterile neutrino
  Dark Matter in gauge extensions of the Standard Model}},''
  \href{http://dx.doi.org/10.1103/PhysRevD.81.085032}{Phys.Rev. {\bf D81}
  (2010)  085032},
\href{http://arxiv.org/abs/0912.4415}{{\tt arXiv:0912.4415 [hep-ph]}}.

\bibitem{Nemevsek:2012cd}
M.~Nemev{\v s}ek, G.~Senjanovi{\'c}, and Y.~Zhang, ``{\em {Warm Dark Matter in
  Low Scale Left-Right Theory}},''
  \href{http://dx.doi.org/10.1088/1475-7516/2012/07/006}{JCAP {\bf 1207} (2012)
   006},
\href{http://arxiv.org/abs/1205.0844}{{\tt arXiv:1205.0844 [hep-ph]}}.

\bibitem{Bonn:2007su}
J.~Bonn, K.~Eitel, F.~Gl{\"u}ck, D.~Sevilla-Sanchez, and N.~Titov, ``{\em {The
  KATRIN sensitivity to the neutrino mass and to right-handed currents in beta
  decay}},'' \href{http://dx.doi.org/10.1016/j.physletb.2011.08.005}{Phys.Lett.
  {\bf B703} (2011)  310--312},
\href{http://arxiv.org/abs/0704.3930}{{\tt arXiv:0704.3930 [hep-ph]}}.

\bibitem{Stephenson:2000mw}
G.~J. Stephenson, J.~T. Goldman, and B.~McKellar, ``{\em {Tritium beta decay,
  neutrino mass matrices and interactions beyond the standard model}},''
  \href{http://dx.doi.org/10.1103/PhysRevD.62.093013}{Phys.Rev. {\bf D62}
  (2000)  093013},
\href{http://arxiv.org/abs/hep-ph/0006095}{{\tt arXiv:hep-ph/0006095
  [hep-ph]}}.

\bibitem{SejersenRiis:2011sj}
A.~Sejersen~Riis, S.~Hannestad, and C.~Weinheimer, ``{\em {Analysis of
  simulated data for the KArlsruhe TRItium Neutrino experiment using Bayesian
  inference}},'' \href{http://dx.doi.org/10.1103/PhysRevC.84.045503}{Phys.Rev.
  {\bf C84} (2011)  045503},
\href{http://arxiv.org/abs/1105.6005}{{\tt arXiv:1105.6005 [nucl-ex]}}.

\bibitem{Barry:2013xxa}
J.~Barry and W.~Rodejohann, ``{\em {Lepton number and flavour violation in
  TeV-scale left-right symmetric theories with large left-right mixing}},''
  \href{http://dx.doi.org/10.1007/JHEP09(2013)153}{JHEP {\bf 1309} (2013)
  153},
\href{http://arxiv.org/abs/1303.6324}{{\tt arXiv:1303.6324 [hep-ph]}}.

\bibitem{Nemevsek:2012iq}
M.~Nemev{\v s}ek, G.~Senjanovi{\'c}, and V.~Tello, ``{\em {Left-Right Symmetry:
  from Majorana to Dirac}},''
  \href{http://dx.doi.org/10.1103/PhysRevLett.110.151802}{Phys. Rev. Lett. {\bf
  110} (2013)  151802},
\href{http://arxiv.org/abs/1211.2837}{{\tt arXiv:1211.2837 [hep-ph]}}.

\bibitem{Farzan:2001cj}
Y.~Farzan, O.~Peres, and A.~Y. Smirnov, ``{\em {Neutrino mass spectrum and
  future beta decay experiments}},''
  \href{http://dx.doi.org/10.1016/S0550-3213(01)00361-3}{Nucl.Phys. {\bf B612}
  (2001)  59--97},
\href{http://arxiv.org/abs/hep-ph/0105105}{{\tt arXiv:hep-ph/0105105
  [hep-ph]}}.

\bibitem{Farzan:2002zq}
Y.~Farzan and A.~Y. Smirnov, ``{\em {On the effective mass of the electron
  neutrino in beta decay}},''
  \href{http://dx.doi.org/10.1016/S0370-2693(03)00207-7}{Phys.Lett. {\bf B557}
  (2003)  224--232},
\href{http://arxiv.org/abs/hep-ph/0211341}{{\tt arXiv:hep-ph/0211341
  [hep-ph]}}.

\bibitem{Ade:2013zuv}
{\bf Planck Collaboration}, P.~Ade {\em et al.}, ``{\em {Planck 2013 results.
  XVI. Cosmological parameters}},''
\href{http://arxiv.org/abs/1303.5076}{{\tt arXiv:1303.5076 [astro-ph.CO]}}.

\bibitem{Boyarsky:2009ix}
A.~Boyarsky, O.~Ruchayskiy, and M.~Shaposhnikov, ``{\em {The Role of sterile
  neutrinos in cosmology and astrophysics}},''
  \href{http://dx.doi.org/10.1146/annurev.nucl.010909.083654}{Ann.Rev.Nucl.Par%
t.Sci. {\bf 59} (2009)  191--214},
\href{http://arxiv.org/abs/0901.0011}{{\tt arXiv:0901.0011 [hep-ph]}}.

\bibitem{Drewes:2013gca}
M.~Drewes, ``{\em {The Phenomenology of Right Handed Neutrinos}},''
  \href{http://dx.doi.org/10.1142/S0218301313300191}{Int.J.Mod.Phys. {\bf E22}
  (2013)  1330019},
\href{http://arxiv.org/abs/1303.6912}{{\tt arXiv:1303.6912 [hep-ph]}}.

\bibitem{Tremaine:1979we}
S.~Tremaine and J.~Gunn, ``{\em {Dynamical Role of Light Neutral Leptons in
  Cosmology}},''
\href{http://dx.doi.org/10.1103/PhysRevLett.42.407}{Phys.Rev.Lett. {\bf 42}
  (1979)  407--410}.

\bibitem{Bulbul:2014sua}
E.~Bulbul, M.~Markevitch, A.~Foster, R.~K. Smith, M.~Loewenstein, {\em et al.},
  ``{\em {Detection of An Unidentified Emission Line in the Stacked X-ray
  spectrum of Galaxy Clusters}},''
  \href{http://dx.doi.org/10.1088/0004-637X/789/1/13}{Astrophys.J. {\bf 789}
  (2014)  13},
\href{http://arxiv.org/abs/1402.2301}{{\tt arXiv:1402.2301 [astro-ph.CO]}}.

\bibitem{Boyarsky:2014jta}
A.~Boyarsky, O.~Ruchayskiy, D.~Iakubovskyi, and J.~Franse, ``{\em {An
  unidentified line in X-ray spectra of the Andromeda galaxy and Perseus galaxy
  cluster}},''
\href{http://arxiv.org/abs/1402.4119}{{\tt arXiv:1402.4119 [astro-ph.CO]}}.

\bibitem{Barbieri:1988av}
R.~Barbieri and R.~N. Mohapatra, ``{\em {Limits on right-handed interactions
  from SN1987A observations}},''
\href{http://dx.doi.org/10.1103/PhysRevD.39.1229}{Phys. Rev. {\bf D39} (1989)
  1229}.

\bibitem{Raffelt:2011nc}
G.~G. Raffelt and S.~Zhou, ``{\em {Supernova bound on keV-mass sterile
  neutrinos reexamined}},''
  \href{http://dx.doi.org/10.1103/PhysRevD.83.093014}{Phys.Rev. {\bf D83}
  (2011)  093014},
\href{http://arxiv.org/abs/1102.5124}{{\tt arXiv:1102.5124 [hep-ph]}}.

\bibitem{Nieves:1986uk}
J.~Nieves, D.~Chang, and P.~Pal, ``{\em {Electric dipole moment of the electron
  in left-right symmetric theories}},''
\href{http://dx.doi.org/10.1103/PhysRevD.33.3324}{Phys. Rev. {\bf D33} (1986)
  3324--3328}.

\bibitem{Chatrchyan:2012meb}
{\bf CMS Collaboration}, S.~Chatrchyan {\em et al.}, ``{\em {Search for
  leptonic decays of $W'$ bosons in $pp$ collisions at $\sqrt{s}=7$ TeV}},''
  \href{http://dx.doi.org/10.1007/JHEP08(2012)023}{JHEP {\bf 1208} (2012)
  023},
\href{http://arxiv.org/abs/1204.4764}{{\tt arXiv:1204.4764 [hep-ex]}}.

\bibitem{atlasnote}
{\bf ATLAS Collaboration}, ``{\em {Search for high-mass states with one lepton
  plus missing transverse momentum in $p p$ collisions at $\sqrt{s} = 8$~TeV
  with the ATLAS detector}}.'' {ATLAS-CONF-2014-017}, 2014.

\bibitem{CMS:2013rca}
{\bf CMS Collaboration}, ``{\em {Search for leptonic decays of $W'$ bosons in
  $p p$ collisions at $\sqrt{s}=8$~TeV}}.'' {CMS-PAS-EXO-12-060}, 2013.

\bibitem{Langacker:1989xa}
P.~Langacker and S.~U. Sankar, ``{\em {Bounds on the Mass of $W_R$ and the
  $W_L$--$W_R$ Mixing Angle $\xi$ in General $SU(2)_L \times SU(2)_R \times
  U(1)$ Models}},''
\href{http://dx.doi.org/10.1103/PhysRevD.40.1569}{Phys. Rev. {\bf D40} (1989)
  1569--1585}.

\bibitem{Huang:2013kma}
W.-C. Huang and J.~Lopez-Pavon, ``{\em {On neutrinoless double beta decay in
  the minimal left-right symmetric model}},''
  \href{http://dx.doi.org/10.1140/epjc/s10052-014-2853-z}{Eur.Phys.J. {\bf C74}
  (2014)  2853},
\href{http://arxiv.org/abs/1310.0265}{{\tt arXiv:1310.0265 [hep-ph]}}.

\bibitem{inder}
P.~Bhupal~Dev, S.~Goswami, M.~Mitra, and W.~Rodejohann, ``{\em {Constraining
  Neutrino Mass from Neutrinoless Double Beta Decay}},''
  \href{http://dx.doi.org/10.1103/PhysRevD.88.091301}{Phys.Rev. {\bf D88}
  (2013)  091301},
\href{http://arxiv.org/abs/1305.0056}{{\tt arXiv:1305.0056 [hep-ph]}}.

\bibitem{Rodejohann:2011mu}
W.~Rodejohann, ``{\em {Neutrino-less Double Beta Decay and Particle
  Physics}},'' \href{http://dx.doi.org/10.1142/S0218301311020186}{Int. J. Mod.
  Phys. {\bf E20} (2011)  1833--1930},
\href{http://arxiv.org/abs/1106.1334}{{\tt arXiv:1106.1334 [hep-ph]}}.

\bibitem{Agostini:2013mzu}
{\bf GERDA Collaboration}, M.~Agostini {\em et al.}, ``{\em {Results on
  neutrinoless double beta decay of 76Ge from GERDA Phase~I}},''
  \href{http://dx.doi.org/10.1103/PhysRevLett.111.122503}{Phys.Rev.Lett. {\bf
  111} (2013)  122503},
\href{http://arxiv.org/abs/1307.4720}{{\tt arXiv:1307.4720 [nucl-ex]}}.

\end{thebibliography}\endgroup

\end{document}